\documentclass[12pt]{article}

\usepackage{epsfig}

\title{DEUTERON BREAK-UP: THEORETICAL DESCRIPTION OF
 POLARIZATION OBSERVABLES IN STAPP FORMALISM} 
\author{O.G.Grebenyuk \footnote{e-mail: olegreb@mail.desy.de}\\ 
{\it Petersburg Nuclear Physics Institute, 188350, Gatchina, Russia}
}
 \date{}

\begin{document}

\maketitle

\begin{abstract}
The vector  $A_y$ and tensor  $A_{yy}$ analyzing powers as well as 
the polarization of the outgoing proton are calculated for the exclusive
deuteron break-up reaction $\vec{d}p \rightarrow \vec{p}pn$ 
at a deuteron beam energy of 2 GeV.
Two component covariant formalism of Stapp has been used to have
the completely Lorentz invariant model.  In addition to the Impulse Approximation
 the nucleon-nucleon double-scattering and delta-excitation
mechanism have been added coherently.
Good agreement  with the precise data obtained at Saclay is achieved .
\\
\\
{\it PACS:} 21.45+v, 24.70+s,  25.10+s\\
\\
\\
{\it Keywords:} NUCLEAR REACTION $\vec{d}(p,\vec{p},p)n$
E=2.0 GeV,  calculated vector analyzing power $A_y$, 
tensor analyzing power $A_{yy}$, polarization of the proton, 
covariant formalism, double-scattering, delta-excitation
\end{abstract}

\clearpage 

\large

\section*{Introduction}

The exclusive break-up of the polarized deuteron by the
protons at a deuteron beam energy of 2 GeV was studied at Saclay 
 \cite{Ero1994,Exp145}. The experiment was motivated by
the idea to explore the important details of the deuteron wave function  
at the high internal momenta.  As compared with
the series of previous $dp\rightarrow ppn$ 
experiments \cite{Perdrisat1969,Witten1975,Felder1976,Perdrisat1985,Sulimov1994},
this experiment was designed to be sensitive to the ratio
between the $S$- and $D$-components of the deuteron wave function brought by 
polarization observables. The sensitivity to the S/D ratio is maximum 
when the Impulse Approximation (IA) is valid. In this approximation one  
of the deuteron's nucleon is a spectator and another one interacts 
with the beam/target proton. 
As theory predicts, and it is confirmed experimentally,  
the IA is valid  only when the internal momentum of 
the deuteron does not exceed  $\approx$ 200 MeV/c.  An attempt of straightforward 
experimental testing of  the deuteron wave function above this momentum could be 
justified only if other mechanisms beyond of the  IA are well under the control. 
Therefore  a good knowledge 
of other interaction mechanisms is of the great  importance in the  
experiments  of such kind.

That is why the theory of 
the correction to the IA has a history as long as the experiments initiated 
by this approximation. Chew and Goldberger 
\cite{ChewGoldberger1952} represented the 
scattering of the elementary particles by complex nuclei as the
multiple-scattering series
$T=\sum_{k=1}^{\infty}T^{(k)}$, 
where $k$ is the number of two-body NN interactions. The $T^{(1)}$
corresponds to the single-scattering and contains the IA term,
$T^{(2)}$ corresponds to the double-scattering and so on.
Analogous development follows from the Faddeev 
equations \cite{Faddeev1961}. Everett 
\cite{Everett1962} was the first who scrutinized 
the double-scattering terms. He used the spin-dependent NN amplitude 
from the phase shift analysis (PSA) corrected so as to allow one nucleon to be 
off-shell and  calculated the loop integrals of the 
double-scattering graphs, taking the 
NN amplitudes out of the integral at a fixed point.  Golovin et al.  
 \cite{Golovin1972} used the similar method with the up-date NN amplitudes.

Another approach was developed by Wallace 
\cite{Wallace1972}. He took an
advantage of the Glauber cancellation by the high-order 
terms in multiple-scattering series of those pieces of the double-scattering 
loop integral, which originate from off-shell states \cite{Harrington1969}. 
This approach obtained an application by Punjabi 
et al. \cite{PerdrisatPunjabi1990}.

The question about the role of the $\Delta$-isobar in the (p,2p)-reactions 
often arises when 
discussing the discrepancies between the experiment  data and 
predictions with the  
only NN rescatterings. Yano \cite{Yano1985} 
calculated the contribution of the $\Delta$-isobar to the deuteron break-up 
reaction 
using  Feynman diagram approach but the corresponding 
amplitude was added  incoherently  to the other amplitudes.
 
The analysis of the copious and precise data on the polarization 
parameters in exclusive $\vec{d}p\rightarrow \vec{p}pn$ reaction,  
obtained at Saclay  \cite{Exp145},  requires to deal 
with the interference between the various amplitudes of the 
multiple scattering. In this paper we present such a model. It is 
based  on the approach of Everett and  includes the spin and isospin variables.
The model takes into account   
the double-scattering contribution and also the graphs including the 
process $\pi^* d\rightarrow$NN, the virtual pion being 
emitted by the beam/target proton.  Since the $\pi d\rightarrow$NN 
reaction at these energies goes mainly through $\Delta$N in the 
intermediate state,  the graphs considered by Yano are also added 
in a coherently to the other contributions. Furthermore, 
the pion-nucleon scattering in other partial waves (S, P, D) is also 
considered.

When calculating the nucleon-nucleon rescattering contributions to the deuteron
break-up amplitude 
one needs to transform the NN matrices from nucleon-nucleon center of mass frames 
to the common laboratory frame. 
These transformations depend on the spin basis chosen for the 
one-nucleon states.  Usually the canonical or helicity basis are
used which are transformed with unitary two by two matrices depending on a 
nucleon momentum. They are so called Wigner rotations and one needs to
make the four different rotations corresponding to the two incoming and
two outgoing scattering nucleons. The covariant basis \cite{Joss1962,Stapp1962}
 transforming independently of the particle
momentum by  the  unimodular two by two  matrices  is
free from this deficiency. An amplitudes in the covariant basis are
usually referred to as the M-functions of Stapp. 
The covariant formalism of the M-functions
developed by Stapp \cite{Stapp1962,Stapp1983} is based on the matrices
 $\sigma^\mu=(1, \vec{\sigma})$ and  $\tilde{\sigma}^\mu=(1, -\vec{\sigma})$, 
where $ \vec{\sigma}$ are the standard Pauli matrices. 
With an each kinematical momentum 
 $P^\mu$ of the reaction  two by two matrices 
$\tilde{P} \equiv P^\mu \tilde{ \sigma_\mu}$ and $P \equiv P^\mu  \sigma_\mu$
are associated. The  products $\tilde{P}_i P_j$ of these matrices
are the  elements from which the M-functions are built.
The matrices  $\tilde{V}_i$ and $V_i$ associated with the
four-velocity of the i-th particle, serve as the metric tensors  
of the particle  when performing the contraction over 
this index (traces, successive processes). 

The outline of the paper is as follows. The Sections 1-3 cover the three
main mechanisms involved in the deuteron break-up: Impulse Approximation,
NN double-scattering and $\Delta$-excitation. The polarization observables 
obtained with the covariant deuteron density matrix are described  
in the \mbox{Section 4.}
The discussion of the calculation results and summary is given in the Section 5.
The introduction to the covariant formalism of the M-functions 
is given in the Appendix.

\vspace{1cm}
\section{Impulse Approximation}

\begin{figure}[ht]
\begin{center}
\begin{picture}(350,170)(0,0)
\thicklines
\put(30,50){\line(1,0){170}} \put(210,50)
{$p_2$, $\tau _2, \sigma _2$}
\put(300,50){$(RS)$}  
\put(20,40){$p_0=(m,\vec{0})$, $\tau _0, \sigma _0$}
\put(30,135){\line(1,0){170}} \put(210,135){$p_3$, $\tau _3, \sigma _3$}
\put(280,135){$(undetected)$}  \put(20,145){$p_d$, $\sigma _p \sigma _n$}
\put(30,130){\line(1,0){50}} \put(80,130){\line(1,-2){40}}
\put(120,50){\line(2,1){80}} \put(210,90){$p_1$, $\tau _1, \sigma _1$}
\put(50,90){$p_v$, $\tau _v, \sigma _v$}
\put(290,80){$(SPES)$}
\end{picture}
\end{center}
\caption{IA mechanism. In Saclay experiment $p_1$ was
measured by the magnetic spectrometer
SPES-4 in coincidence with the recoil protons ($p_2$) detected 
by Recoil Spectrometer RS.}
\label{fig:IA}
\end{figure}
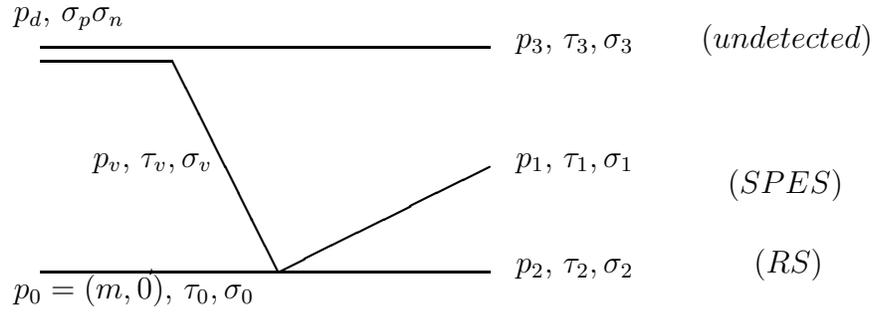

The general expression for the $S$-matrix elements of the deuteron break-up
reaction is 
\[S=i(2 \pi )^4 \delta ^4(p_1+p_2+p_3-p_0-p_d) M \;,\]
where $M$ is the amplitude of the reaction and 
we assign indices for the
each particle of the reaction d(p,2p)n as
\begin{center}
\begin{tabular}{cccccccccl}
   d  & + &  p  & $\rightarrow$ &  p  & + &  p  & + &  n  & \\
  (d) & + & (0) & $\rightarrow$ & (1) & + & (2) & + & (3) &. \\
\end{tabular}
\end{center}
To avoid in what follows the numerous repetitions of the summation
sign $\sum$, which is inevitable when considering the spin and
isotopic spin variables, we will use  the tensor-like notation with
the indices characterizing spin and isotopic spin projections. The
upper and lower indices  relate to the final and initial channels,  
respectively. The same index at the up and down positions stands
for a summation over this index.
Free nucleon  is characterized  by the projections of the spin
$\sigma$, of the isospin $\tau$  and by the momentum $p$.
We will use the  representation of the deuteron spin states by the
symmetric spin-tensor $S^{\sigma _n \sigma _p}$ so that the amplitudes
of the reaction with participation of the deuteron will have the pair
of  indices $\sigma _n \sigma _p$ and will be symmetric with respect to
their interchange. 

The amplitude of  the deuteron break-up reaction  with the all indices looks like 
\[{{\rm M}^{\tau _1 \tau _2 \tau _3}_{\tau _0}}^
{; \sigma _1 \sigma _2 \sigma _3}_
{; \sigma _0 \sigma _p \sigma _n}(p_1,p_2,p_3;p_0,p_d)\;.\] 
 The Pauli principle requires  that this amplitude should be antisymmetric
with respect to the interchange of any two final nucleons. For example
\[{{\rm M}^{\tau _1 \tau _2 \tau _3}_{\tau _0}}^
{; \sigma _1 \sigma _2 \sigma _3}_
{; \sigma _0 \sigma _p \sigma _n}(p_1,p_2,p_3;p_0,p_d)= 
-{{\rm M}^{\tau _2 \tau _1 \tau _3}_{\tau _0}}^
{; \sigma _2 \sigma _1 \sigma _3}_
{; \sigma _0 \sigma _p \sigma _n}(p_2,p_1,p_3;p_0,p_d)\;. \]
To fulfill this requirement we  proceed as follows. Having written
the expression antisymmetric with respect to the interchange of for
example the first and the second nucleons ${\rm M}^{(12)3}$, we
perform then the cyclic permutation of the nucleons. The resulting
expression ${\rm M}^{(12)3}+{\rm M}^{(23)1}+{\rm M}^{(31)2}$ 
possess the required antisymmetry.

Let us start with the expression  corresponding to the graph pictured
in  Fig.  \ref{fig:IA}. With the all  indices it looks like 
\begin{equation}
{{\rm M}_{NN}}_{\tau _0 \tau _v; \sigma _0 \sigma _v}
^{\tau _1 \tau _2; \sigma _1 \sigma _2}(p_1,p_2;p_0,p_v)\; 
\frac{i}{2m_v(m_v-m)}\; 
{{\rm D}^{\tau _v \tau _3;}}^{\sigma _v \sigma _3}_
{\sigma _p \sigma _n}(p_v,p_3;p_d)\;, 
\label{eq:ONE d-3}
\end{equation}
where $m_v^2 \equiv (p_d-p_3)^2$ is the squared mass of the virtual nucleon
and ${\rm M}_{NN}$ and ${\rm D}$ are the  nucleon-nucleon amplitude 
and the deuteron vertex, respectively. The above expression corresponds only 
to the first term in the  development of the spinor particle 
propagator in the Dirac particle-antiparticle formalism
\[\frac{\not p +m}{s-m^2}=\frac{1}{2W}\left[ \frac{u \cdot \bar{u}}{W-m} - 
\frac{v \cdot \bar{v}}{W+m}\right]\;,\;\;\;W=\sqrt{p^2}\;.\]
Thus we do not consider in what follows the virtual antinucleon states, 
which requires a knowledge of the antinucleon  components of the nucleon-nucleon
amplitude and of the deuteron vertex.
The deuteron wave function  is a \mbox{product} of the vertex and the propagator
\begin{equation}
\frac{{{\rm D}^{\tau _v \tau _3;}}^{\sigma_v \sigma_3}_{\sigma_p \sigma_n}
(p_v,p_3;p_d)}
{2m_v(m_v-m)}  \equiv \epsilon ^{\tau _v \tau _3}
\Phi^{\sigma_v \sigma_3}_{\sigma_p \sigma_n}(p_v,p_3;p_d)\;,
\label{eq:dnp wave funtion}
\end{equation}
where $ \epsilon$ is an antisymmetric two by two tensor in isotopic space. 
The fully  antisymmetrized 'one nucleon exchange' ($ONE$) amplitude is equal to
\begin{eqnarray}
{{{\rm M}_{ONE}}^{\tau _1 \tau _2 \tau _3}_{\tau _0}}^
{; \sigma _1 \sigma _2 \sigma _3}_
{; \sigma _0 \sigma _p \sigma _n}(p_1,p_2,p_3;p_0,p_d)= \nonumber \\
i[{{\rm M}_{NN}}_{\tau _0 \tau _v; \sigma _0 \sigma _v}
^{\tau _1 \tau _2; \sigma _1 \sigma _2}(p_1,p_2;p_0,p_d-p_3)
\epsilon ^{\tau _v \tau _3} 
\Phi^{\sigma _v \sigma _3}_{\sigma _p \sigma _n}(p_d-p_3,p_3;p_d)+ \nonumber \\
{{\rm M}_{NN}}_{\tau _0 \tau _v; \sigma _0 \sigma _v}
^{\tau _2 \tau _3; \sigma _2 \sigma _3}(p_2,p_3;p_0,p_d-p_1)
\epsilon ^{\tau _v \tau _1} 
\Phi^{\sigma _v \sigma _1}_{\sigma _p \sigma _n}(p_d-p_1,p_1;p_d)+
\label{eq:ONE term}  \\
{{\rm M}_{NN}}_{\tau _0 \tau _v; \sigma _0 \sigma _v}
^{\tau _3 \tau _1; \sigma _3 \sigma _1}(p_3,p_1;p_0,p_d-p_2)
\epsilon ^{\tau _v \tau _2} 
\Phi^{\sigma _v \sigma _2}_{\sigma _p \sigma _n}
(p_d-p_2,p_2;p_d)]\;, \nonumber
\end{eqnarray} 
where we have taken into account the symmetry properties of 
nucleon-nucleon amplitude in the expression (\ref{eq:ONE d-3}).

The IA stands for the deuteron break-up amplitude represented  
by only  the  term in the 
eq.(\ref{eq:ONE term}) with the minimal virtuality of the intermediate nucleon.
It corresponds to  the minimal momentum of nucleon spectator in the 
deuteron at rest frame. Under the kinematical conditions 
of the Saclay experiment it was 
the undetected neutron ($p_3$).

In the next subsections we describe the nucleon-nucleon amplitude 
and the deuteron wave function.

\vspace{1cm}
\subsection{NN amplitude}

\begin{figure}[h]
\begin{center}
\begin{picture}(200,100)(0,0)
\thicklines
\put(100,50){\circle{40}}
\put(30,30){\line(1,0){140}} \put(30,70){\line(1,0){140}}
\put(50,20){$p_0$, $\tau_0$, $\sigma_0$}
\put(50,80){$p_v$, $\tau_v$, $\sigma_v$}
\put(160,20){$p_2$, $\tau_2$, $\sigma_2$}
\put(160,80){$p_1$, $\tau_1$, $\sigma_1$}
\end{picture}
\end{center}
\caption{NN amplitude}
\label{fig:NN amplitude}
\end{figure}
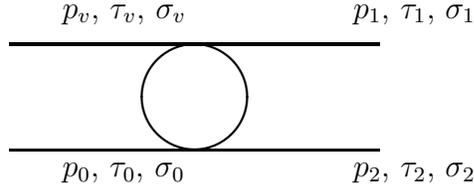
 The spin and isospin dependent NN amplitude looks like
\begin{eqnarray*}
&{{\rm M}_{NN}}^{\tau _1,\tau _2 ;\sigma _1\sigma _2}
_{\tau _0,\tau _v ;\sigma _0 \sigma _v}
(p_1,p_2;p_0,p_v) =&    \\
& {\displaystyle \frac{e^{\tau _1}_{\tau _0}e^{\tau _2}_{\tau _v}-
e^{\tau _2}_{\tau _0}e^{\tau _1}_{\tau _v}}{2} }
{{\rm M}_{0}}^{\sigma _1\sigma _2}_{\sigma _0\sigma _v}(p_1,p_2;p_0,p_v)+
{\displaystyle \frac{e^{\tau _1}_{\tau _0}e^{\tau _2}_{\tau _v}+
e^{\tau _2}_{\tau _0}e^{\tau _1}_{\tau _v}}{2} }
{{\rm M}_{1}}^{\sigma _1\sigma _2}_{\sigma _0\sigma _v}(p_1,p_2;p_0,p_v)\;,&
\end{eqnarray*}
where $e$ is the unit two by two operator in
the isotopic space and  the M$_0$ and  M$_1$ are  the isosinglet and
 isotriplet  parts of the NN amplitude, respectively.
According to the Pauli principle the  amplitudes M$_0$ and  M$_1$
obey the  symmetry relations
\begin{eqnarray*}
{{\rm M}_{0}}^{\sigma _1 \sigma _2}_{\sigma _0 \sigma _v}(p_1,p_2;p_0,p_v)=
{{\rm M}_{0}}^{\sigma _2 \sigma _1}_{\sigma _0 \sigma _v}(p_2,p_1;p_0,p_v)\;,
\nonumber \\
{{\rm M}_{1}}^{\sigma _1 \sigma _2}_{\sigma _0 \sigma _v}(p_1,p_2;p_0,p_v)=
-{{\rm M}_{1}}^{\sigma _2 \sigma _1}_{\sigma _0 \sigma _v}(p_2,p_1;p_0,p_v)\;.
\end{eqnarray*}
The c.m. canonical amplitudes, i.e. the
$S$-matrix elements, of the NN scattering are expressed  in terms of five 
complex amplitudes $a,b,c,d,e$ \cite{Bystritsky1978}  
\[S=\frac{1}{2}[(a+b)+(a-b)\hat{n} \otimes \hat{n}
+(c+d)\hat{m} \otimes \hat{m}
+(c-d)\hat{l} \otimes \hat{l}+
e(e \otimes \hat{n}+\hat{n} \otimes e)]\;,\]
where   
\[\vec{m}=\frac{\vec{k}_f-\vec{k}_i}{|\vec{k}_f-\vec{k}_i|}\;,\;\;
\vec{l}=\frac{\vec{k}_f+\vec{k}_i}{|\vec{k}_f+\vec{k}_i|}\;,\;\;
\vec{n}=\frac{\vec{k}_i \times \vec{k}_f}{|\vec{k}_i \times \vec{k}_f|}\;,\]
 $\vec{k}_{i,f}$ are the initial and final c.m. momenta 
of the NN interacting system  and 
$\hat{n} \equiv \vec{n} \cdot \vec{\sigma}$ and so on. The shorthand
$A \otimes B$ stands for $A^{\sigma_1}_{\sigma_v}B^{\sigma_2}_{\sigma_0}$.

We have to deal with the NN amplitudes in the laboratory frame and  it is 
inconvenient to make the transformations from numerous individual NN c.m. frames
to these frame during the calculations. 
It is why the applying of the M-functions of 
Stapp \cite{Stapp1962,Stapp1983} is very natural in such calculations.
The basic M-functions  analogous to the
$S$-matrix basis $e \otimes e, \hat{m} \otimes \hat{m}, 
\hat{l} \otimes \hat{l}, \hat{n} \otimes \hat{n},
e \otimes \hat{n}+\hat{n} \otimes e$ are built from the 
products  $\tilde{V}_iV_j$ of the two by two (hermitian) matrices 
$\tilde{V}_i \equiv V_i^\mu \tilde{ \sigma_\mu}$ and 
$V_j \equiv V_j^\mu  \sigma_\mu$, where  $V^\mu _i$ are the
four-velocities of the scattered nucleons completed by the 
four-velocity $V$ of the whole system in the arbitrary frame. 
The possible M-function's basis $b_i,\;i=1, \ldots ,6$ is
given by the eqs.(\ref{eq:bi}-\ref{eq:norms Mi}) of the Appendix. Once fixed, 
the basis allows represent the M-function  of the NN scattering as a sum
\begin{equation}
{\rm M}=g_1b_1+g_2b_2+g_3b_3+g_4b_4+g_5\frac{b_5+b_6}{\sqrt{2}}\;,
\label{eq:M func devel}
\end{equation}
where $g_i$ are the five complex amplitudes.
The relations between the M-function's amplitudes $g_i$ and 
the canonical amplitudes 
$a,b,c,d,e$ are \cite{Grebenyuk1989}
\begin{eqnarray*}
g_1=-\frac{c+d}{2}\;,& g_2={\displaystyle -\frac{c-d}{2} }\;,   \\
g_3=\frac{b+a\cos \varphi -ie\sin \varphi }{2}\;, &
g_4= {\displaystyle \frac{b-a\cos \varphi +ie\sin \varphi }{2} }\;,  \\
g_5=-\frac{a\sin \varphi +ie\cos \varphi }{\sqrt{2}}\;,&
e^{i\varphi}= {\displaystyle \frac{\omega_0 -\omega}{\omega_0 -\omega^{-1}} }\;, 
\end{eqnarray*}
where $\omega =e^{i\theta}$, $\theta$ being the c.m. scattering angle,  and 
\[\omega_0 \equiv \sqrt{\frac{(V,V_2)+1}{(V,V_2)-1}\;
\frac{(V,V_0)+1}{(V,V_0)-1}}\;.\]
The values of the amplitudes {\em a, b, c, d, e} for given laboratory energy   
and c.m. angle  were calculated by use of the PSA of Arndt et 
al. \cite{Arndt1988}. The normalization of the M-functions
is such that the c.m. cross-section is equal to 
\[\frac {d \sigma }{d \Omega_{NN} }=\frac {1}{{(8 \pi )}^2s} \;
\frac{Tr({\rm M}\;\;\tilde{V}_v \otimes \tilde{V}_0\;\;
{\rm M}^\dagger \;\;V_1 \otimes V_2)}{4}\;.\]
The presence of metric matrices $\tilde{V}_i$ and $V_j$ is the peculiarity
 of the Stapp formalism as was 
mentioned in the Introduction. 

As to the virtual nucleons they are off-shell and we took it into account
only kinematically. In particular we calculate the basis M-functions $b_i$ 
by use of the eqs.(\ref{eq:bi}-\ref{eq:norms Mi}) using the 'virtual' velocity
$V_v=(p_d-p_3)/m_v$, if of course
\mbox{$m_v^2= (p_d-p_3)^2$} was happened to be positive.  When it was not 
 the case  we assigned 
to the matrix elements of the corresponding nucleon-nucleon amplitude  
 zero values. 
For  the scalar amplitudes $g_i$ the  only way to obtain the off-shell values is 
the model calculations and they are permanently in progress 
(see for example \cite{Gross1992}).
Still the situation is far from being satisfactory, especially 
for the energies above several hundreds MeV. 
Thus only  the on-shell amplitudes obtained in the PSA
are of the practical use.  The choice of the on-shell kinematic 
corresponding to the  off-shell one  is ambiguous.  One  way is to take 
the PSA solution at the
 $s=(p_0+p_v)^2=(p_1+p_2)^2$ and $t=(p_1-p_0)^2=(p_2-p_v)^2$. It means to take 
$g_i(s,t,m_v^2) = g_i(s,t,m^2)$. Another way
is to put at first the virtual nucleon on the mass-shell 
$p_v=(E_v,{\bf p}_v) \rightarrow p_v^*=(\sqrt{{\bf p}_v^2+m^2},
{\bf p}_v)$, and then get the $PSA$ solution at the
$s^*={(p_0+p_v^*)}^2 \geq s$ and $t^*={(p_2-p^*_v)}^2$, i.e. to accept that
$g_i(s,t,m_v^2) = g_i(s^*,t^*,m^2)$.
We preferred the first way, since it ignores the 
off-shell mass dependence at all in contrast to the non-dynamical recipe  for such
dependence induced by the second way.

\vspace{1cm} 
\subsection{Deuteron wave function}

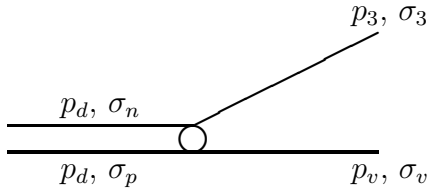
\begin{figure}[ht]
\begin{center}
\begin{picture}(200,100)(0,0)
\thicklines
\put(100,25){\circle{10}}
\put(30,20){\line(1,0){140}} \put(30,30){\line(1,0){70}}
\put(100,30){\line(2,1){70}}
%\put(140,17){X}
\put(50,10){$p_d$, $\sigma _p$}
\put(50,35){$p_d$, $\sigma _n$}
\put(160,10){$p_v$,  $\sigma _v$}
\put(160,70){$p_3$,  $\sigma _3$}
\end{picture}
\end{center}
\caption{DNP-vertex}
\label{fig:dnp vertex}
\end{figure}

The DNN-vertex (see Fig.  \ref{fig:dnp vertex})  relates to 
the wave function by the eq.(\ref{eq:dnp wave funtion}). The antisymmetric tensor
\[\epsilon ^{\tau _v \tau _3}=\frac{1}{\sqrt{2}}
\left( \begin{array}{cc}
0 & 1 \\
-1 & 0
\end{array}\right) \; \]
determines the isosinglet nature of the deuteron. 
The classic deuteron wave functions are related to the deuteron at  rest frame.
The most suitable for the transition to the M-function
expression looks like 
\[{\Phi_s} ^{\sigma_v \sigma_3}_{\sigma_p \sigma_n}(\vec{k})=
a( e^{\sigma_v}_{\sigma_p}
 e^{\sigma_3}_{\sigma_n} +
 e^{\sigma_v}_{\sigma_n}
 e^{\sigma_3}_{\sigma_p})  -b k^ik^j
( {e_i}^{\sigma_v}_{\sigma_p}
 {e_j}^{\sigma_3}_{\sigma_n} +
 {e_i}^{\sigma_v}_{\sigma_n}
 {e_j}^{\sigma_3}_{\sigma_p})\;,  \]
where subscript $s$ reminds the $S$-matrix origin of the classic wave function
and the unit vector $\vec{k}$ is directed along the nucleon momentum. The
matrices $e_i,\;i=1,2,3$ coincide  with the Pauli matrices, but we use the
different notation to distinguish them from the matrices $\sigma^i$, 
which have another spinor indices:
$\sigma^i_{\bar{c}d}$. The scalar functions $a$ and $b$ are connected with
the $S$- and $D$-wave functions  $u$ and $w$ as following
\begin{eqnarray}
a=u-\frac{w}{\sqrt{8}}\;\;,&\;\;
b=-{\displaystyle \frac{3w}{\sqrt{8}}}\;, \label{eq:a,b(u,w)} \\
u=a-\frac{b}{3}\;\;,&\;\;w=-{\displaystyle \frac{b\sqrt{8}}{3}}\;. \nonumber
\end{eqnarray}
The corresponding M-function of the deuteron wave function in an arbitrary 
frame is built from the 
products $\tilde{V}_vV_d$ and $\tilde{V}_3V_d$ of the two by two matrices,                      
 where $V^\mu_{d,v,3}$ are the four-velocities of the
deuteron, virtual nucleon and on-shell nucleon, respectively. It is equal to 
\begin{eqnarray}
\Phi^{\sigma_v \sigma_3}_{\sigma_p \sigma_n}(p_v,p_3;p_d)=
\nonumber \\  a\frac{
 (e+\tilde{V_v}V_d)^{\sigma_v}_{\sigma_p}
 (e+\tilde{V_3}V_d)^{\sigma_3}_{\sigma_n} +
 (e+\tilde{V_v}V_d)^{\sigma_v}_{\sigma_n}
 (e+\tilde{V_3}V_d)^{\sigma_3}_{\sigma_p}}
{2\sqrt{((V_v,V_d)+1)((V_3,V_d)+1)}}
 + \nonumber \\  b\frac{
 (e-\tilde{V_v}V_d)^{\sigma_v}_{\sigma_p}
 (e-\tilde{V_3}V_d)^{\sigma_3}_{\sigma_n} +
 (e-\tilde{V_v}V_d)^{\sigma_v}_{\sigma_n}
 (e-\tilde{V_3}V_d)^{\sigma_3}_{\sigma_p}}
{2\sqrt{((V_v,V_d)-1)((V_3,V_d)-1)}}\;.
\label{eq:M matrix triplet vertex}
\end{eqnarray}
The following normalization equation holds
\[Tr(\Phi \tilde{\rho}_0 \Phi^{\dagger}\;V_3 \otimes V_v) =
\Phi^{\sigma _v \sigma _3}_{\sigma _p \sigma _n}
\rho_0^{\sigma_p \bar{\sigma}_p\;\;\sigma_n \bar{\sigma}_n}
\bar{\Phi}^{\bar{\sigma} _v \bar{\sigma} _3}_
{\bar{\sigma} _p \bar{\sigma}_n}
{V_3}_{\bar{\sigma}_3 \sigma_3}{V_v}_{\bar{\sigma}_v \sigma_v}=u^2+w^2\;,\]
where $\tilde{\rho}_0$ is the invariant density matrix of the deuteron 
defined in the Section 4 (see eq.(\ref{eq:r0})).
The functions $u$ and $w$ obtained with the Paris \cite{Paris} 
or Bonn \cite{Bonn} potentials
are approximated by series of the poles
\begin{eqnarray}
u(m_d^2,m_v^2)=\sqrt{8\pi m_d}N_S \left(\frac{1}{q^2+\alpha^2}
-\sum_i\frac{c_i}{q^2+\alpha_i^2}\right)\;,
\nonumber \\
w(m_d^2,m_v^2)=\sqrt{8\pi m_d}N_D \left(\frac{1}{q^2+\alpha^2}
-\sum_i\frac{d_i}{q^2+\alpha_i^2}\right)\;.
\label{eq:u,w param}
\end{eqnarray}
The values $N_{S,D}$ are the normalization constants ($N_S^2 \simeq 0.16\;GeV$)
and the dependence on the $m_d^2$ and $m_v^2$ goes through the 3-momentum 
of the nucleons in the deuteron at rest frame
\[q=\frac{\sqrt{[(m_d+m)^2-m_v^2][(m_d-m)^2-m_v^2]}}{2m_d}\;.\] 
The binding   energy $\epsilon$ relates  to the $\alpha$ as $\alpha^2=\epsilon m$.
The sum rules $\sum_i  c_i= \sum_i d_i=1$ should be fulfilled. 
The dimension of the wave functions $u$ and $w$ is  
GeV$^{-1}$ and they are normalized as follows
\[\int d \vec{q} (u^2+w^2)=(2\pi)^3 2m_d\;.\]  

\vspace{1cm}
\section{ NN double-scattering}

\begin{figure}[ht]
\begin{center}
\begin{picture}(250,100)(0,0)
\thicklines 
\put(80,75){\circle{10}} 
\put(30,80){\line(1,0){170}} \put(30,70){\line(1,0){50}}
\put(80,70){\line(1,-1){42}}
\put(30,20){\line(1,0){170}}
\put(125,25){\circle{10}} 
\put(170,75){\circle{10}} 
\put(170,70){\line(2,-1){30}} 
\put(128,28){\line(1,1){42}}
\put(30,90){$p_d$, $\sigma _p \sigma _n$}
\put(30,10){$p_0$, $\tau _0, \sigma _0$}
\put(210,10){$p_2$, $\tau _2, \sigma _1$}
\put(210,55){$p_1$, $\tau _1, \sigma _1$}
\put(210,90){$p_3$, $\tau _3, \sigma _3$}
\put(100,90){$p_s$, $\tau _s, \sigma _s$}
\put(30,45){$p_v$, $\tau _v, \sigma _v$}
\put(150,45){$p_f$, $\tau _f, \sigma _f$}
\put(125,77){X}
\end{picture}
\end{center}
\caption{Double-scattering mechanism}                                 
\label{fig:DS}
\end{figure}
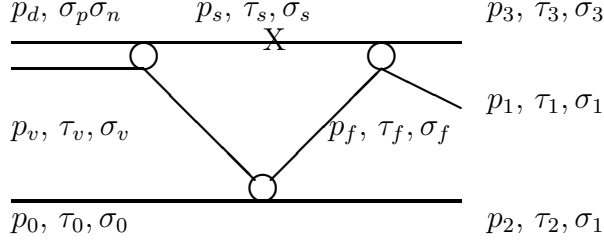

Let us start with the  graph shown in Fig. \ref{fig:DS} referred to 
in what follows as the
DS graph 2(31). Two other graphs with  the cyclically permuted nucleons will be 
referred to as the DS graphs 1(23) and 3(12). For the DS graph 2(31) the 
amplitude with the  'spectator' on the mass-shell looks like
\begin{eqnarray}
&{{{\rm M}^{2(31)}_{DS}}^{\tau _1 \tau _2 \tau _3}_{\tau _0}}^
{;\sigma _1 \sigma _2 \sigma _3}_
{;\sigma _0 \sigma _p \sigma _n}(p_1,p_2,p_3;p_0,p_d)= 
-i \int {\displaystyle\frac{d{\bf p}_ s}{(2\pi)^3 2E_s}} & \nonumber \\
&{\rm M}^{\tau _3 \tau _1;\sigma _3 \sigma _1}_
{\tau _f \tau _s;\sigma _f \sigma _s}(p_3,p_1;p_s,p_{31}-p_s) 
{\rm M}^{\tau _2 \tau _f;\sigma _2 \sigma _f}_
{\tau _0 \tau _v;\sigma _0 \sigma _v}(p_2,p_{31}-p_s;p_0,p_d-p_s)&
\label{eq:DS loop int} \\
&{\displaystyle \frac{\epsilon ^{\tau _v \tau _s} \Phi^{\sigma _v \sigma _s}_
{\sigma _p \sigma _n}(p_d-p_s,p_s;p_d)}{2m_f(m_f-m+i\varepsilon)}}&\nonumber
\end{eqnarray}
where \mbox{$p_{31} =p_3+p_1=(E_{31},{\bf p}_{31})$}, 
$E_s=\sqrt{m^2+{\bf p}_ s^2}$ and $m_f^2=(p_{31}-p_s)^2$.
The simplifications are necessary to
calculate of the integral (\ref{eq:DS loop int}) since it requires 
a knowledge of the 
off-shell nucleon-nucleon amplitudes. 
The most simple method is to take the nucleon-nucleon amplitudes  out of the 
integral sign. Reasonable to do it at some momentum  
$p_s^0$ placed on the singular 
surface of the integral  corresponding to the mass-shell 
of the virtual nucleon $f$.
With the $z$-axis  directed  along the momentum ${\bf p}_{31}$ the equation of 
this  surface looks like
\begin{equation} 
(\frac{p^x_s}{q_{31}})^2+ (\frac{p^y_s}{q_{31}})^2+ 
(\frac{p^z_s-\frac{|{\bf p}_{31}|}{2}}{q_{31}\frac{E_{31}}{W_{31}}})^2=1\;,
\label{eq:ellipse}
\end{equation} 
where $W_{31}=\sqrt{s_{31}}=\sqrt{p^2_{31}}$ is the invariant mass of 
the 31-pair of the nucleons and $q_{31}=\sqrt{s_{31}-4m^2}/2$ is their c.m.
momentum. At the end of this Section and in the Section 5 presenting the 
calculation results we return to the problem of the  
choice of this Fermi momentum. 

When the nucleon pair at the final state  has the invariant mass 
near the nucleon-nucleon threshold  it is necessary and possible to take 
into account the off-shell behavior of the corresponding
NN amplitude in the eq.(\ref{eq:DS loop int})  to
fulfill  the closure sum rules in the $^1 S_3$  state
(see an argumentation  in 
\cite{Aladashvili1977,AlvearWilkin1984,KolybasovKsenzov1975}). 
Near the threshold it is possible to describe this off-shell behavior 
by the simple form-factor ${\rm M}^{off}={\rm M}^{on}\;f(s_{31},m_f^2)$, which 
is related to the $^1S_3$ wave function of the deuteron (\ref{eq:u,w param}) as 
follows
\begin{equation}
f(s_{31},m_f^2)=1-(q^2+\alpha^2)\sum_i\frac{c_i}{q^2+\alpha_i^2}\;,
\label{eq:off-shell ff} 
\end{equation}
where the dependence on the virtual mass and energy goes from the c.m. momentum 
\[q=\frac{\sqrt{[(W_{31}+m)^2-m^2_f][(W_{31}-m)^2-m^2_f]}}{2W_{31}}\;.\] 
We used the form-factor (\ref {eq:off-shell ff}) up to the
energy 200 MeV and above this energy we were replacing it by the unity.
Note that at the low energies  the corresponding DS graph is refereed to as 
the final state interaction (FSI) graph. 

Taking the NN amplitudes 
out of the integral sign,  but keeping inside of it the off-shell form-factor,
deuteron wave function and the propagator, we derive the following approximation
of the DS amplitude from the eq.(\ref{eq:DS loop int}) 
\begin{eqnarray}
&{{{\rm M}^{2(31)}_{DS}}^{\tau _1 \tau _2 \tau _3}_{\tau _0}}^
{;\sigma _1 \sigma _2 \sigma _3}_
{;\sigma _0 \sigma _p \sigma _n}(p_1,p_2,p_3;p_0,p_d)\simeq &\nonumber \\
&{\displaystyle -i\epsilon ^{\tau _v \tau _s} 
{\rm M}^{\tau _3 \tau _1;\sigma _3 \sigma _1}_
{\tau _f \tau _s;\sigma _f \sigma _s}(p_3,p_1;p^0_s,p_{31}-p^0_s)
{\rm M}^{\tau _2 \tau _f;\sigma _2 \sigma _f}_
{\tau _0 \tau _v;\sigma _0 \sigma _v}(p_2,p_{31}-p^0_s;p_0,p_d-p^0_s)}&
\nonumber   \\
& {\displaystyle F^{\sigma _v \sigma _s}_
{\sigma _p \sigma _n}(s_{31},t_2)}\;,& \label{eq:DS31}
\end{eqnarray}
where 
\[F^{\sigma _v \sigma _s}_
{\sigma _p \sigma _n}(s_{31},t_2) \equiv                            
\int \frac{d{\bf p}_ s}{(2\pi)^3 2E_s}
\frac{\Phi^{\sigma _v \sigma _s}_
{\sigma _p \sigma _n}(p_d-p_s,p_s;p_d)f(s_{31},(p_{31}-p_s)^2)}
{2m_f(m_f-m+i\varepsilon)}\]
and $t_2=(p_2-p_0)^2=(p_d-p_{31})^2$.
The deuteron wave function $\Phi$ is defined by the 
eq.(\ref{eq:M matrix triplet vertex}).
To preserve the covariance it is necessary to approximate 
the velocities of the virtual nucleon ($v$) and  the nucleon 'spectator' ($s$)
 in the latter integral by the corresponding values  in 
the NN amplitudes taken out of the integral sign. 
Then the expression for the $F$ becomes
\begin{eqnarray*}
& F^{\sigma _v \sigma _s}_
{\sigma _p \sigma _n}(s_{31},t_2) \simeq 
 {\displaystyle \frac{F_a}{2}\frac{
 (e+\tilde{V_v}V_d)^{\sigma_v}_{\sigma_p}
 (e+\tilde{V_s}V_d)^{\sigma_s}_{\sigma_n} +
 (e+\tilde{V_v}V_d)^{\sigma_v}_{\sigma_n}
 (e+\tilde{V_s}V_d)^{\sigma_s}_{\sigma_p}}
{2\sqrt{((V_v,V_d)+1)((V_s,V_d)+1)}}}
  + & \\ & {\displaystyle \frac{F_b}{2}\frac{
 (e-\tilde{V_v}V_d)^{\sigma_v}_{\sigma_p}
 (e-\tilde{V_s}V_d)^{\sigma_s}_{\sigma_n} +
 (e-\tilde{V_v}V_d)^{\sigma_v}_{\sigma_n}
 (e-\tilde{V_s}V_d)^{\sigma_s}_{\sigma_p}}
{2\sqrt{((V_v,V_d)-1)((V_s,V_d)-1)}}}\;,&
\end{eqnarray*}
where
\[V_v=\frac{p_d-p^0_s}{\sqrt{(p_d-p^0_s)^2}}\;,\;\;
V_s=\frac{p^0_s}{m}\;,\;\;V_d=\frac{p_d}{m_d} \]
are the four-velocities of the corresponding particles.
The complex functions $F_{a,b}$ are equal to (see eqs.(\ref{eq:a,b(u,w)}))
\[F_a=F_u-\frac{F_w}{\sqrt{8}}\;\;,\;\;
F_b=-\frac{3F_w}{\sqrt{8}}\;,  \]
where
\[F_u (s_{31},t_2)\equiv \int \frac{d{\bf p}_ s}{(2\pi)^3 2E_s}
\frac{u(m_d^2,(p_d-p_s)^2)f(s_{31},(p_{31}-p_s)^2) }
{2m_f(m_f-m+i\varepsilon)}\;\]
and $F_w$ is determined analogously by the $w$-wave function. For
the pole expansion of the functions $u$ and $w$ (\ref{eq:u,w param})
a good analytical approximation of these integrals have been obtained 
 by J.M.Laget \cite{Laget1978}:
\begin{eqnarray} 
&{\displaystyle F_u \simeq \frac {N_S} { q_{31}\sqrt{32 \pi m_d}} \{ }&\nonumber \\
&{\displaystyle 2\left[\arctan \frac{p_+}{\alpha }+ \arctan \frac{p_-}{\alpha }
-\sum_i  c_i \left(\arctan \frac{p_+}{\alpha_i}+ 
\arctan \frac{p_-}{\alpha_i} \right) \right]} & \nonumber \\
&{\displaystyle +4\left[\sum_i c_i \; 
\arctan \frac{q_{31}}{2(\alpha + \alpha _i)}- \;
\sum _{i,j}c_i c_j \; \arctan \frac{q_{31}}{2(\alpha_i + \alpha _j)}\right]- }& 
\label{eq:Lag appr}\\
& {\displaystyle i\left[\ln \frac{p_-^2+\alpha ^2}{p_+^2+\alpha ^2}-
\sum _ic_i\ln \frac{p_-^2+ \alpha _i^2}{p_+^2+ \alpha _i^2}\right] \} }\;,&
\nonumber
\end{eqnarray}
where \[q_{31}=\frac{\sqrt{[(W_{31}+m_d)^2-t_2][(W_{31}-m_d)^2-t_2]}}{2m_d}\]
is the momentum of 31-pair in the deuteron at rest frame and
 \[p_{\pm } \equiv \frac{q_{31}}{2} \pm \frac{s_{31}+m_d^2-t_2}{2W_{31}m_d}
\; \frac{\sqrt{s_{31}-4m^2}}{2}\;.\]  
The replacements in the eq.(\ref{eq:Lag appr}) 
of the normalization constant $N_S$ by the $N_D$ 
and of the parameters $c_i$ by the $d_i$ 
entail the expression for the  $F_w$. We have tried 
in the calculations the $F_w$ thus 
obtained  in addition to the $F_u$, but in the final calculations 
it was still omitted. The reason for 
this lays in the closure sum rule, which requires in case of  the $F_w \not =0$  
to take into account the off-shell behavior of the FSI nucleon-nucleon 
amplitude also in the $^1D_3$-state, rather than in the $^1S_3$-state only as 
we did.

The real part
of the factor $F$ in the eq.(\ref{eq:Lag appr}) has two terms. The first one
is due to the off-shell states contribution inside the triangle loop, 
which is proved to be canceled in the eikonal regime \cite{Harrington1969}. 
The second term in the real part (\ref{eq:Lag appr}) 
originates from the threshold form factor
(\ref{eq:off-shell ff}) and  we were dropping it for the  energies
of the finally interacting nucleons above 200 MeV.  

Now let us return to the problem of the choice of the Fermi momentum $p_s^0$ in 
the eq.(\ref{eq:DS31}).
We have considered two candidates. The first one is 
advocated as follows. The  deuteron wave function in the integral 
(\ref{eq:DS loop int}) drops quickly with the decreasing of the virtual mass
$m_v^2=(p_d-p_s)^2$  suppressing by this the contribution of
${\bf p}_s$ corresponding to the small virtual masses $m_v$.
It is reasonable then to test the $p_s^0$ giving the maximum value of $m_v$.  In 
the c.m. frame of the 31-pair the momenta placed on the ellipsoidal surface
(\ref{eq:ellipse}) look like $(\frac{W_{13}}{2},q_{31}{\bf n})$, where
${\bf n}$ is an unite vector. Then
\[m_v^2=(p_d-p^0_s)^2=m_d^2+m^2-E_d W_{13}+2p_d q_{31} x\;,\] 
where $(E_d,{\bf p}_d)$ is the deuteron four-momentum in this frame and  $x$ 
is the cosine of the angle between ${\bf n}$  and  ${\bf p}_d$.  
Therefore the maximum $m_v^2$ is achieved at $x=1$ and the 
momentum of the nucleon $s$  should be directed along the deuteron momentum
in the 31-pair c.m. frame. We will denote the momentum 
thus obtained as $p_s^{max}$.
The argumentation for the second candidate is less evident and consists of
the following. Since the states with small virtual masses are still contribute
to the integral (\ref{eq:DS loop int}) we should not use the 
Fermi momentum corresponding to the maximal $m_v$  but try a momentum corresponding
to some smaller value of the virtual mass  $m_v$. We have chosen for this purpose 
the momentum placed at the 'top' of the ellipsoidal surface (\ref{eq:ellipse}),
which corresponds to the maximum velocity of the nucleon $s$ 
in the laboratory frame.
We will call this momentum as the optimal one and will denote 
it as $p_s^{opt}$. With the $z$-axis  directed  
along the momentum ${\bf p}_{31}$ its components are
\[p_s^{opt}=(E_s,0,0, \frac{|{\bf p}_{13}|}{2}+
q_{31}\frac{E_{13}}{W_{13}} )\;,\]
whereas the  momentum $p_s^{max}$ has in this frame the components
\[p_s^{max}=(E_s,q_{31}n_x,q_{31}n_y, \frac{|{\bf p}_{13}|}{2}+
q_{31}\frac{E_{13}}{W_{13}} n_z)\;,\]
where ${\bf n}$ is the unite vector directed along 
the deuteron momentum in c.m. frame of the 31-pair.

The comparison with the experiment  of the calculations 
performed using the both Fermi momenta 
is presented in the  Section 5 and definitely testifies 
in favor of the optimal Fermi momentum $p_s^{opt}$.

\vspace{1cm}
\section{$\Delta$ -excitation contribution}

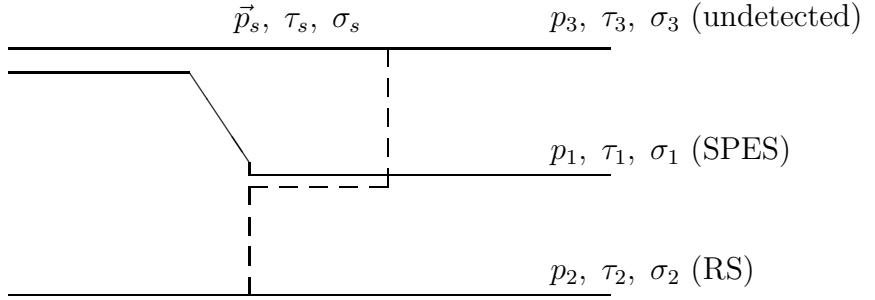
\begin{figure}[t]
\unitlength=.80mm
\begin{center}
\begin{picture}(114.00,50.00)(0,95)
\put(14.00,139.00) {\line(1,0){100}}
\put(14.00,135.00) {\line(1,0){30}}
\put(44.00,135.00) {\line(2,-3){10}}
\put(54.00,118.00) {\line(1,0){60}}  
\put(14.00, 98.00) {\line(1,0){100}}   
\put(54.00, 98.00) {\line(0,3){3}}  
\put(54.00,103.00) {\line(0,3){3}}   
\put(54.00,108.00) {\line(0,3){3}} 
\put(54.00,113.00) {\line(0,3){3}}  
\put(54.00,118.00) {\line(0,3){2}} 
\put(54.00,116.00) {\line(3,0){3}}  
\put(59.00,116.00) {\line(3,0){3}}      
\put(64.00,116.00) {\line(3,0){3}}      
\put(69.00,116.00) {\line(3,0){3}}      
\put(74.00,116.00) {\line(3,0){3}}    
\put(77.00,116.00) {\line(0,3){3}}     
\put(77.00,121.00) {\line(0,3){3}}    
\put(77.00,126.00) {\line(0,3){3}}    
\put(77.00,131.00) {\line(0,3){3}}     
\put(77.00,136.00) {\line(0,3){3}}   
\put(62.00,144.00){\makebox(0,0)[cc]{$\vec{p}_s,\;\tau_s,\;\sigma_s$}}
\put(104.00,144.00){\makebox(0,0)[lc]{$p_3,\;\tau_3,\;\sigma_3$     (undetected)}}
\put(104.00,102.00){\makebox(0,0)[lc]{$p_2,\;\tau_2,\;\sigma_2$     (RS)}}
\put(104.00,122.00){\makebox(0,0)[lc]{$p_1,\;\tau_1,\;\sigma_1$     (SPES)}}
\end{picture}
\end{center}
\caption{$\Delta$ excitation mechanism}
\label{fig:Delta excitation}
\end{figure}
The contribution corresponding to the graph shown in 
Fig.  \ref{fig:Delta excitation} is equal to
\begin{eqnarray}
{{{\rm M}^{2(31)}_{\Delta}}^{\tau _1 \tau _2 \tau _3}_{\tau _0}}^
{;\sigma _1 \sigma _2 \sigma _3}_
{;\sigma _0 \sigma _p \sigma _n}(p_1,p_2,p_3;p_0,p_d)= \nonumber \\
i\frac{{\Gamma ^t}^{\tau _2 ;\sigma_2}_{\tau _0;\sigma _0}(p_2,q;p_0)
{{\rm M}^{\tau _1 \tau _3}_t}^{;\sigma _1 \sigma _3}_
{;\sigma _p \sigma _n}(p_1,p_3;q,p_d)}{q^2-\mu^2}
\;,\;\;\;\;q=p_0-p_2\;,
\label{eq:delta contr}
\end{eqnarray}
where the $\mu$ is the pion mass.
The $\pi$NN-vertex in the eq.(\ref{eq:delta contr}) is equal to
\[{\Gamma ^t}^{\tau _2 ;\sigma _2}_{\tau _0; \sigma _0}
(p_2,q;p_0) ={e^t}^{\tau _2}_{\tau _0}
G^{\sigma _2}_{\sigma _0}(p_2,q;p_0)\;,\]
where $t$ is the isovector index of the pion and 
\[e^{+1}=\left(\begin{array}{cc}
0 &  0 \\
1 &  0
\end{array}\right)\;, \;\;
e^{-1}=\left(\begin{array}{cc}
0 &  1 \\
0 &  0
\end{array}\right)\;, \;\;
e^0=\frac{1}{\sqrt{2}}\left(\begin{array}{cc}
1 &  0 \\
0 & -1
\end{array}\right)\;. \]
The spatial part of $\pi$NN-vertex with the virtual pion looks in Stapp 
formalism like (see the eq.(\ref{eq:indless pseuds vertex}) of the Appendix)
\[G^{\sigma _2}_{\sigma _0}(p_2,q;p_0)=f(q^2) g_\pi
m (e-\tilde{V}_2 V_0)^{\sigma_2}_{\sigma_0}\;,\]
where $g_{\pi} \simeq 13.6$ is the $\pi$NN coupling constant, and
$f(q^2)$ is the pion form factor, for which we took the monopole
representation
\begin{equation}
f(q^2) =\frac{m_\pi ^2-\Lambda
^2}{q^2-\Lambda ^2}
\label{eq:piNN ff}
\end{equation}
with the cut-off $\Lambda =1.0$ GeV.

The  spin and isospin dependent $\pi d\rightarrow$NN  
amplitude in the eq.(\ref{eq:delta contr}) looks like
\[{{\rm M}^{\tau _1 \tau _3}_t}^{;\sigma _1 \sigma _3}_
{;\sigma _p \sigma _n}(p_1,p_3;q,p_d)=
\epsilon ^{\tau _1 \tau _3}_t
M^{\sigma _1 \sigma _3}_{\sigma _p \sigma _n}(p_1,p_3;q,p_d)\;,\]
where
\[\epsilon _{+1}=\left( \begin{array}{cc}
1 & 0 \\
0 & 0
\end{array} \right)\;\; ,\; \;
\epsilon _{-1}=\left( \begin{array}{cc}
0 & 0 \\
0 & -1
\end{array}\right)\;\; ,\; \;
\epsilon _{0}=-\frac{1}{\sqrt{2}}\left( \begin{array}{cc}
 0        & 1 \\
1 & 0
\end{array}\right)\;.\]
The amplitudes of the physical reactions are equal then to
\[{\rm M}_{\pi ^{+}d \rightarrow pp}=M\;\; ,\;\;
{\rm M}_{\pi ^{-}d \rightarrow nn}=-M\;\; ,\;\;
{\rm M}_{\pi ^{0}d \rightarrow np}=-\frac{M}{\sqrt{2}}\;. \]
We have used for the deriving of the M-function of  the $\pi d\rightarrow$NN
reaction the $S$-matrix elements obtained by J.M.Laget \cite{Laget1981}.
In this approach  the one-loop box diagrams with the 
N$\Delta$ as the intermediate state had been calculated numerically, including 
the $\pi$N intermediate scattering in the S, P and D waves  parametrized by 
their phase shifts. However, to avoid the double counting with the DS term 
we have excluded
%%%%%%%% Clarify%%%%%%% 
%the one nucleon exchange diagram  and 
%%%%%%%%%%%%%%%%%%%%%%%
the $P_{11}$ -wave 
(the nucleon pole in the $\pi$N amplitude is a part of the DS term). Finally, 
the $\rho$ -exchange was considered in the  
\mbox{$\pi d\rightarrow$NN} amplitude. The details are given in the Appendix of the
paper \cite{Laget1981}. These calculations have been performed 
in the deuteron at rest  frame. 

So having at our disposal the $S$-matrix elements 
\[S^{\sigma_1\sigma_3}_{\sigma_p\sigma_n}(p_{1(d)},p_{3(d)};q_{(d)},m_d)\;,\]
where the subscript $(d)$ means that the momenta are considered 
in the deuteron at rest 
frame, we have to obtain the M-function in an arbitrary frame.  The only 
way in this case is the straightforward applying  of the common recipe 
prescribed by the eq.(\ref{eq:S=invvfMvi}) of the Appendix. At first we 
derive the M-function in the deuteron at rest system: 
\begin{equation}
M^{\sigma_1\sigma_3}_{\sigma_p\sigma_n}(p_{1(d)},p_{3(d)};q_{(d)},m_d)=
{v_1}^{\sigma_1}_a {v_3}^{\sigma_3}_b 
S^{a\;\;b}_{\sigma_p\sigma_n}(p_{1(d)},p_{3(d)};q_{(d)},m_d)\;.
\label{eq:M in d-rest}
\end{equation}
The boost matrix  $v^a_b$ of the nucleon from the rest to the four-velocity 
$(V_0,{\bf k}\sqrt{V_0^2-1})$, where ${\bf k}$ is the unit vector along the 
momentum, is determined by the eq.(\ref{eq:boost}).
Having obtained the M-matrix in the deuteron at rest frame we applied to the 
eq.(\ref{eq:M in d-rest}) the boost transformation ${v_d}^a_b$ corresponding 
to the deuteron moving along the $z$-axis with the four-velocity 
$(V_{d0},0,0,\sqrt{V_{d0}^2-1})$ so that the final M-function looked like
(see the eq.(\ref{eq:Ml=zMinvz}))
\[M^{\sigma_1\sigma_3}_{\sigma_p\sigma_n}(p_1,p_3;q,p_d)=
{v_d}^{\sigma_1}_a {v_d}^{\sigma_3}_b M^{a\;\;b}_{g\;\;h}
(p_{1(d)},p_{1(d)};q_{(d)},m_d)
{v^{-1}_d}^g_{\sigma_p}{v^{-1}_d}^h_{\sigma_n} \;.\]

\vspace{1cm}
\section{Deuteron density matrix and observables}

Let us remind at first how like looks the density matrix  of a spinor particle  
in the Stapp formalism  \cite{Stapp1983}:
\begin{equation}
\rho^{a\bar{b}}=\frac{1}{2}(V^\mu+S^\mu)\sigma_\mu^{a\bar{b}}\;,
\label{eq:nucl den matr} 
\end{equation}
where the $a,b=1,2$ are the spinor indices, $V^\mu$ is the four-velocity 
and $S^\mu$ is the four-vector of polarization of the particle, which is
orthogonal to $V$: $(V,S)=0$. In the index-less form it looks like
\[\tilde{\rho}=\frac{1}{2}(V^\mu+S^\mu)\tilde{\sigma}_\mu=
\frac{1}{2}(\tilde{V}+\tilde{S})\;.\]

The analog of the expression (\ref{eq:nucl den matr}) for the vector particle is
\begin{equation}
\rho ^ {\sigma_p \bar{\sigma}_p\;\;\sigma_n \bar{\sigma}_n}=
\rho^{\mu \nu} \sigma_\mu^{\sigma_p \bar{\sigma}_p}
\sigma_\nu^{\sigma_n \bar{\sigma}_n}\;,
\label{eq:ro in dir prod}
\end{equation}
where
\[\rho ^ {\mu \nu}=\frac{1}{12}\left[-g^{\mu \nu}+4V^\mu V^\nu +
2\sqrt{3}(S^\mu V^\nu +S^\nu V^\mu)+
4\sqrt{\frac{3}{2}}T^{\mu \nu}\right]\;.\]
The $S$ and $T$ are the vector and tensor polarizations with the properties
\[(S,V)=0\;,\; T^{\mu \nu}=T^{\nu \mu}\;,\; V_\mu T^{\mu \nu}=0\;,\;
 T^\mu _\mu=0\;.\]
In the index-less form eq.(\ref{eq:ro in dir prod}) could be written as
\[\tilde{\rho}=\rho^{\mu \nu} \tilde{\sigma}_\mu \otimes \tilde{\sigma}_\nu\;.
\]
If the quantization axis is directed along the $y$-axis then
in the c.m. frame we have
\begin{eqnarray*}
S=(0,0,\frac{\sqrt{3}}{2}p_y,0)\;, \nonumber \\
T=-\frac{p_{yy}}{2\sqrt{6}}\left(
\begin{array}{cccc}
0 & 0 &  0 & 0 \\
0 & 1 &  0 & 0 \\
0 & 0 & -2 & 0 \\
0 & 0 &  0 & 1
\end{array} \right)\;,
\end{eqnarray*}
where $p_y \equiv n_+-n_-$ and $p_{yy}\equiv n_++n_--2n_0$. The
statistical weights $n_+,n_-,n_0$ of the states 
with the definite projections of the spin 
on the quantization axis are normalized so that 
$n_++n_0+n_-=1$. Then $Tr(\rho^2)=n_+^2+n_0^2+n_-^2$.
If the particle moves along the $z$-axis with the velocity
$V=(V_0,0,0,V_z)$, then $S$ does not change, but $T$ becomes
\[T_{\mu \nu}=-\frac{p_{yy}}{2\sqrt{6}}\left(
\begin{array}{cccc}
V_z^2  & 0 & 0  & V_0V_z   \\
 0     & 1 & 0  & 0        \\
 0     & 0 & -2 & 0        \\
V_0V_z & 0 & 0  & V_0^2
\end{array} \right) \equiv -\frac{p_{yy}}{2\sqrt{6}}t_{\mu \nu}\;.\]
We can represent then the $\tilde{\rho}$  as
\begin{equation}
\tilde{\rho}_d=
\tilde{\rho}_0+\frac{3}{2}p_y\tilde{\rho}_y+\frac{1}{2}p_{yy}\tilde{\rho}_{yy}\;,
\label{eq:r0+ry+ryy}
\end{equation}
where
\begin{equation}
\tilde{\rho}_0 \equiv \frac{1}{3}\tilde{V} \otimes \tilde{V} -
\frac{1}{12}\tilde{\sigma}_\mu \otimes \tilde{\sigma}^\mu
\label{eq:r0}
\end{equation}
is the unpolarized density matrix and 
\begin{eqnarray*}
\tilde{\rho}_y \equiv -\frac{1}{3}(\tilde{\sigma}_y \otimes \tilde{V} +
\tilde{V} \otimes \tilde{\sigma}_y)\;, \\
\tilde{\rho}_{yy} \equiv -\frac{1}{6} t_{\mu \nu} \tilde{\sigma}^\mu \otimes
 \tilde{\sigma}^\nu
\end{eqnarray*}
are the vector and the tensor polarization matrices, respectively. 
If the (target) proton
is unpolarized, the initial density matrix of the $dp$ system is equal to 
\[\tilde{\rho}_i=\frac{\tilde{V}_0}{2} \otimes \tilde{\rho}_d\]
and the vector and tensor analyzing powers are equal  to
\begin{eqnarray}
\sigma_0 A_y=\frac{Tr(M \;\;\tilde{V}_0 \otimes \tilde{\rho}_y\;\;
M^\dagger \;\;V_1 \otimes V_2 \otimes V_3)}{2}\;,
\label{eq:s0ay}    \\
\sigma_0 A_{yy}=\frac{Tr(M \;\;\tilde{V}_0 \otimes \tilde{\rho}_{yy}\;\;
M^\dagger\;\; V_1 \otimes V_2 \otimes V_3)}{2}\;,    
\label{eq:s0ayy}
\end{eqnarray}
where
\[\sigma_0 \equiv \frac{Tr(M \;\;\tilde{V}_0 \otimes \tilde{\rho}_0\;\;
M^\dagger\;\; V_1 \otimes V_2 \otimes V_3)}{2}\; \]
is the unpolarized cross section.
The polarization of the fast outgoing nucleon in the $y$ direction is 
determined by the equation 
\[\sigma P_{1y}=Tr(M \;\; \tilde{\rho}_i\;\; M^\dagger \;\;\sigma^y 
\otimes V_2 \otimes V_3)\;, \]
where the polarized cross section is equal to
\[\sigma =Tr(M  \;\;\tilde{\rho}_i\;\; M^\dagger \;\;V_1 
\otimes V_2 \otimes V_3)=
\sigma_0(1+\frac{3}{2}p_yA_y+\frac{1}{2}p_{yy}A_{yy})\;.\]
The last equation follows from 
the eqs.(\ref{eq:r0+ry+ryy},\ref{eq:s0ay},\ref{eq:s0ayy}).
Introducing the polarization $P_0$ of this nucleon for the unpolarized 
deuteron and the depolarization parameters $D_v$ as follows
\begin{eqnarray*}
\sigma_0 P_0 = \frac{Tr(M \;\;\tilde{V}_0 \otimes \tilde{\rho}_0\;\;
M^\dagger\;\; \sigma^y \otimes V_2 \otimes V_3)}{2}\;,  \\ 
\sigma_0 D_v = \frac{Tr(M \;\;\tilde{V}_0 \otimes \tilde{\rho}_y\;\;
M^\dagger\;\; \sigma^y \otimes V_2 \otimes V_3)}{2} 
\end{eqnarray*}
we can write polarization $P_{1y}$ for the vector
polarized deuteron beam in the following form
\begin{equation}
 P_{1y}=\frac{P_0+\frac{3}{2}p_yD_v}
{1+\frac{3}{2}p_yA_y}\;.
\label{eq:vec pol d  p1y}
\end{equation}

\vspace{1cm}
\section{Calculation results and discussion} 

Let us start with the main parameters of the detecting system of 
the performed experiment.
Polarized deuteron beam of the accelerator Saturne was incident on a 4 cm thick 
liquid hydrogen target.
Fast protons from the reaction ($p_1$) were selected at the  
scattering angle $\Theta_1=18.0 \pm 0.5^\circ$ by the magnetic spectrometer
SPES-4 in the coincidence with the recoil protons ($p_2$) detected at the
angle  $\Theta_2=57.0^\circ ( \pm 4.25^\circ$ in the scattering plane and 
$\pm 8^\circ$ in the vertical plane) with a mosaic of $E$ and $\Delta E$
scintillation counters. Two multiwire proportional chambers  at 1.5 
and 3 $m$ from the target provided the recoil proton track position. Six central
momentum settings  for the proton detected in SPES-4 
($p_{10}=$ 1.6, 1.7, 1.8, 1.9, 2.0, 2.05 GeV/c) were studied. The 
polarimeter POMME located behind the final focal plane of SPES-4 was used to
measure the polarization of the fast protons.
At the chosen  $\Theta_1$ and for the momentum $p_1$ of the fast proton, two
kinematic solutions are possible for the recoil proton energy $T_2$,
which according to our definition are the high energy (HES) and  the low energy  
(LES) solutions.
The neutron spectator momenta in the deuteron rest frame, denoted as $q$, 
ranged from 30 to 440 MeV/c.

Thus the calculation procedure should include  scanning over the acceptance of 
the detecting system. By scanning over the allowed phase space of the detectors,
we intended to obtain a realistic calculation result, which can be directly 
compared to the data. This scanning procedure inevitably invokes the necessity 
for a Monte-Carlo type event generation, because the number of phase unit 
volumes amounts to a huge number when the unit volume is defined by the 
resolution of the measurement.  

\begin{figure}
\begin{center}
\psfig{figure=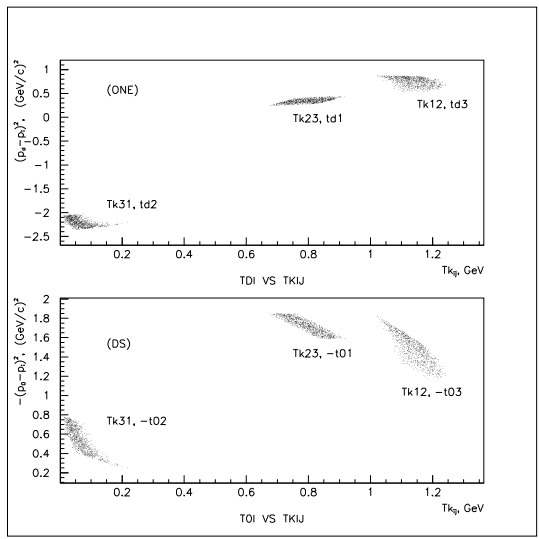,bbllx=20pt,bblly=145pt,bburx=575pt,bbury=700pt,width=7cm}
\end{center}
\caption{The $(T_{ij},t_{di})$ and $(T_{ij},t_{0i})$ distributions
         for $p_{10}=$ 1.6 GeV/c and HES.} 
\label{fig:Tij td0i}
\end{figure}

We would like to show now,  taking the 
$p_{10}=$ 1.6 GeV/c and HES as an example, some kinematic distributions 
characterizing different mechanisms of the deuteron break-up in 
this experiment.                           
In Fig. \ref{fig:Tij td0i} the two-dimensional 
distributions of the kinetic energies 
$T_{ij}=(s_{ij}-4m^2)/
{2m}$ of the $i,j$ pairs of the final nucleons  with two momenta transfer,
$t_{di}=(p_d-p_i)^2$ and $t_{0i}=(p_0-p_i)^2$, are shown. Note that $t_{di}$ are 
equal to the squared masses of virtual  nucleon in the nucleon pole graphs. 
It is seen  that of three final nucleons  
considered as the spectator in these pole graphs, 
the neutron ($p_3$) provides the maximal 
masses of the exchanged nucleon. The pole 
graph with $p_2$ as the spectator gives  the negative squared masses and we 
have excluded this graph from the calculation. The distributions of 
$(T_{ij},t_{0i})$  characterize the graphs of the DS mechanism.
$T_{31}$ ranges from 30 to 200 Mev and therefore the DS graph 2(31) is the 
typical FSI graph. Two other pairs 
of the final nucleons and especially two protons $p_1$ and $p_2$ have much 
higher relative energies. However
applying of the Glauber approach when calculating the DS graphs with these  
pairs being rescattered \cite{Wallace1972,PerdrisatPunjabi1990} is not 
justified because the momenta transfer $t_{03}$ and $t_{01}$ for these graphs 
are very high.
Thus we have applied  for calculation of the DS graphs the expression 
(\ref{eq:DS31}) and two others obtained from it by the cyclic permutation of the
final nucleons. 

\begin{figure}
\begin{center}
\psfig{figure=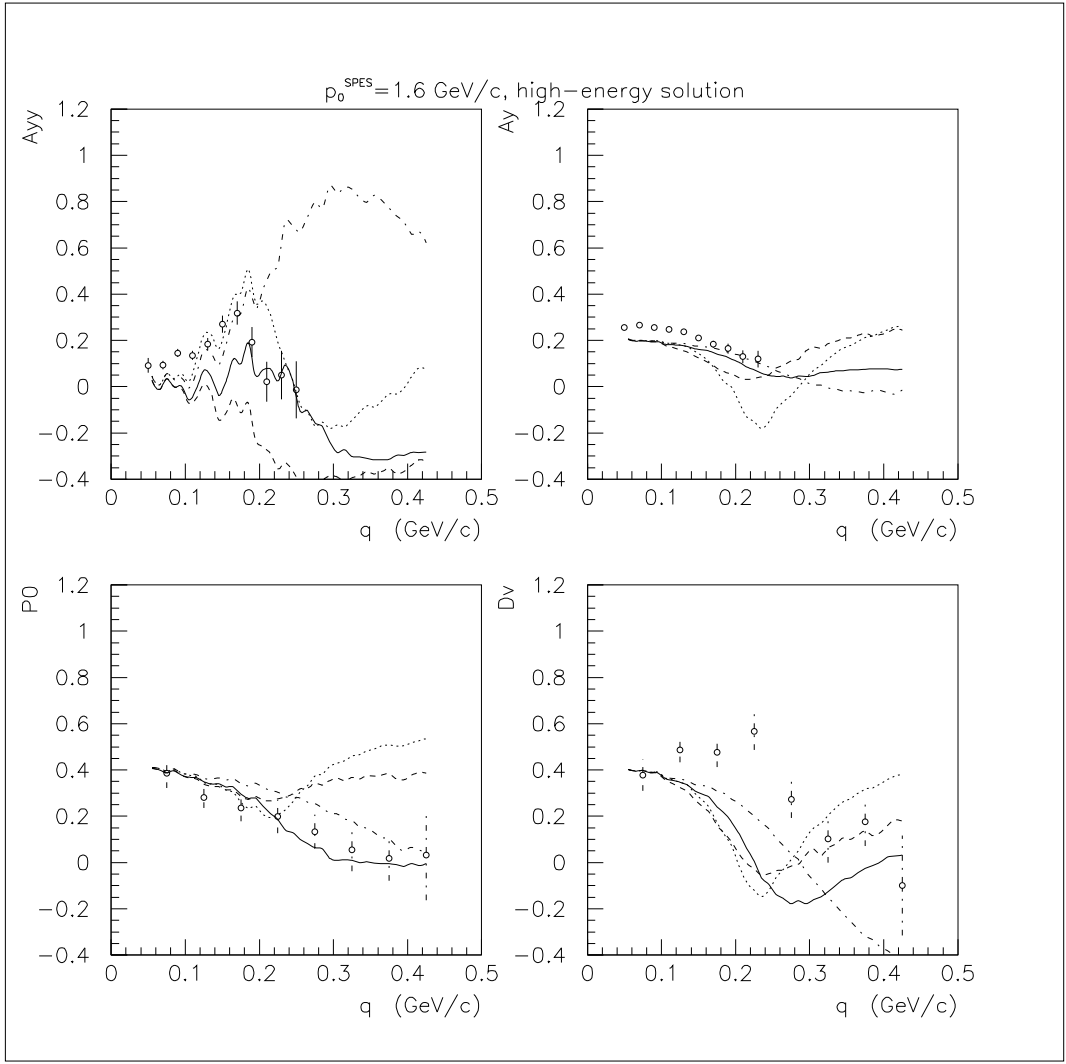,bbllx=20pt,bblly=145pt,bburx=575pt,bbury=700pt,width=14cm}
\end{center}
\caption{Polarization observables for the different choices of the Fermi momentum.
Dotted line  presents the calculations
with  $p^{max}_s$  in the DS graph 1(23) and $p^{opt}_s$  in the DS graph 3(12).
Dashed line 
corresponds to the choice of $p^{opt}_s$ for DS graph 1(23) and $p^{max}_s$  
for the DS graph 3(12). Solid line is obtained with the $p^{opt}_s$ in the both
these graphs. The dashed-dotted line presents 
the only FSI prediction (DS graph 2(31)).} 
\label{fig:com of fm}
\end{figure}

We would like now to return to the  problem mentioned in the
\mbox{Section 2}. It is the choice of the Fermi momentum $p^0_s$
in the eq.(\ref{eq:DS31}). 
The dotted line in Fig. \ref{fig:com of fm} presents 
the polarization observables calculated
with  the $p^{max}_s$  in the DS graph 1(23) and with the $p^{opt}_s$  
in the DS graph 3(12)
(the sensitivity of the contribution of the FSI to the choice of Fermi
momentum is weak and we fixed it to be  $p^{max}_s$). The dashed line 
corresponds to the choice of the $p^{opt}_s$ for the DS graph 1(23) and 
of the $p^{max}_s$  
for the DS graph 3(12). The solid line is obtained with the $p^{opt}_s$ in the both
these graphs. The improvement in the description of the  data in the
latter case is evident. It is important that  neglecting of these graphs
results in the significant worsening of the data description. 
The dashed-dotted line presents the only FSI prediction and 
we see that it does not provide the suppression of the tensor analyzing power
at the high $q$ observed in the experiment. So we can summarize that besides
the FSI graph the both 1(23) and  3(12) DS graphs are required 
for data description and they should
be calculated with $p^{opt}_s$ as the Fermi momentum. It is the very choice
of the DS model, which we have applied in further calculations.

\begin{figure}
\begin{center}
\epsfig{file=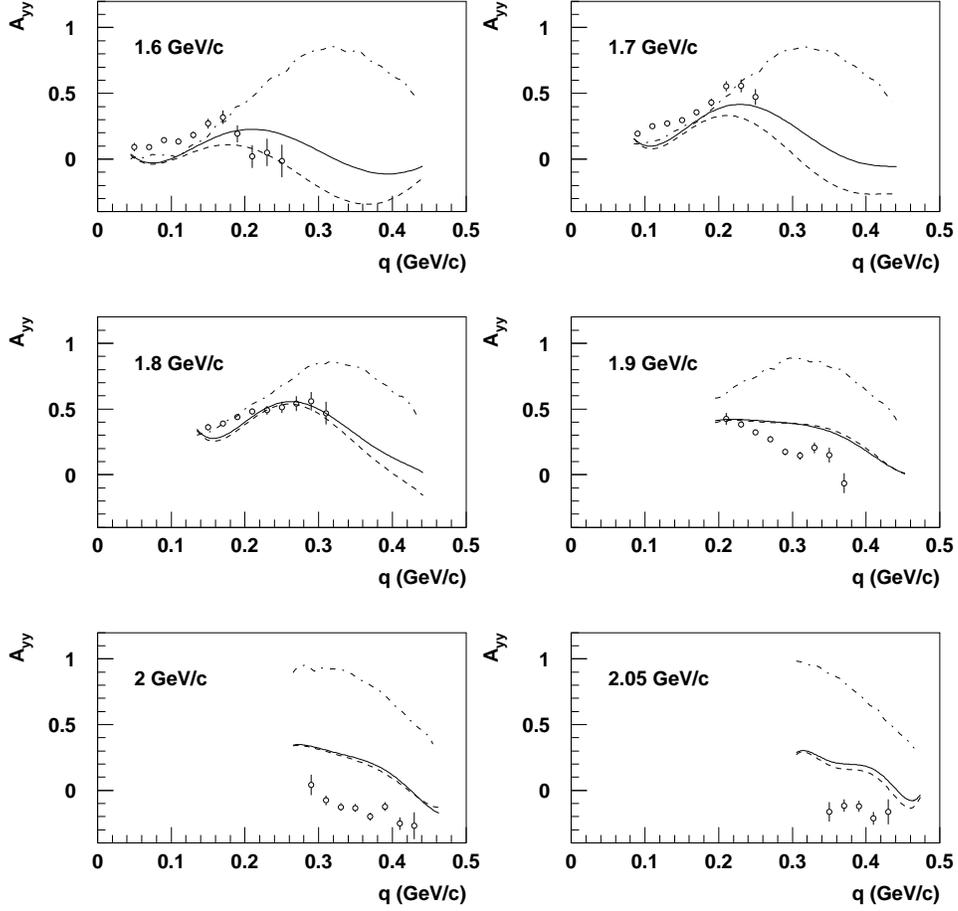,height=14cm}
\end{center}
\caption{The tensor analyzing power $A_{yy}$ for the HES. 
The experimental points are 
presented
for the different values of the central momentum detected in the magnetic 
spectrometer and as a
function of the outgoing neutron momentum in the deuteron at rest frame. 
The dashed-doted line is the IA ,
the dashed line has in addition the DS contribution, and the continuous line is 
the full calculation including in addition the virtual $\Delta$.}
\label{fig:Ayy}
\end{figure}

\begin{figure}
\begin{center}
\epsfig{file=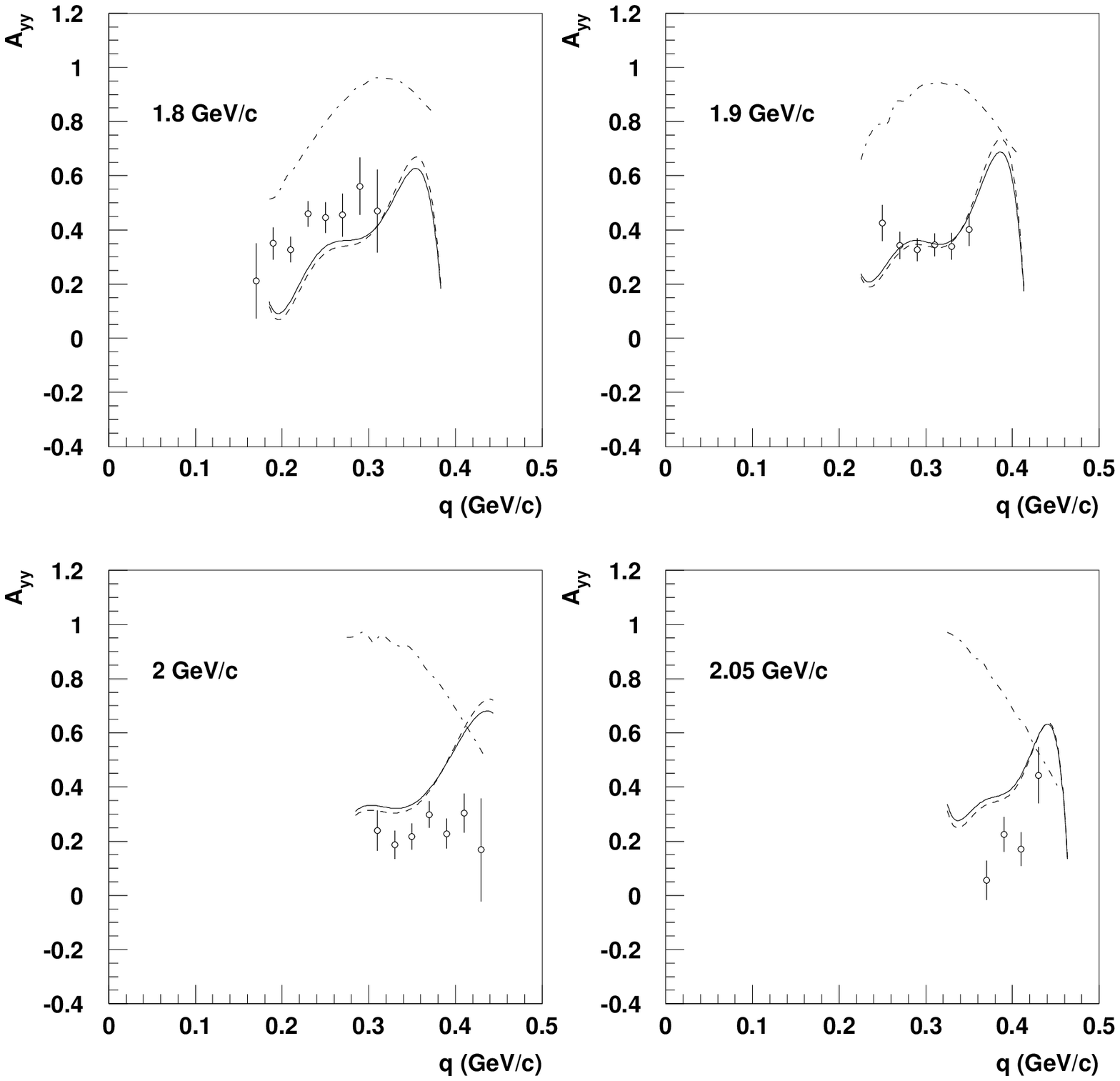,height=16cm}
\end{center}
\caption{The tensor analyzing power $A_{yy}$ for the LES.
 Same notations as in Fig. \ref{fig:Ayy}. }
\label{fig:leAyy}
\end{figure}

The measured tensor analyzing powers $A_{yy}$ and the calculation 
results with Bonn deuteron wave function are shown 
in Figs. \ref{fig:Ayy} and \ref{fig:leAyy}.
The main feature of the experimental data is the strong deviation from
the IA (dashed line) for $q \geq 0.2$ GeV/c. The full theory including
the $\Delta$-excitation graphs (full line) gives the
satisfactory explanation of this deviation, the DS graphs (dashed-dotted
line) playing the decisive role. One could expect  the more noticeable
$\Delta$-effects taking in mind that the laboratory kinetic energies of the
23-pair of the final nucleons range from 0.7 to 0.9 GeV 
(see Fig. \ref{fig:Tij td0i}),
which is exactly the region of the $\Delta$ dominance in the $\pi d\rightarrow$NN
amplitude. Yet the momentum transfer $t_{01}$ from the target proton
to the fast proton detected by the SPES-4 is very high and the pion form
factor (\ref{eq:piNN ff}) reduces the common contribution of the graph
shown in  Fig. \ref{fig:Delta excitation} ( with, of course, correspondingly
interchanged final nucleons).

\begin{figure}
\begin{center}
\epsfig{file=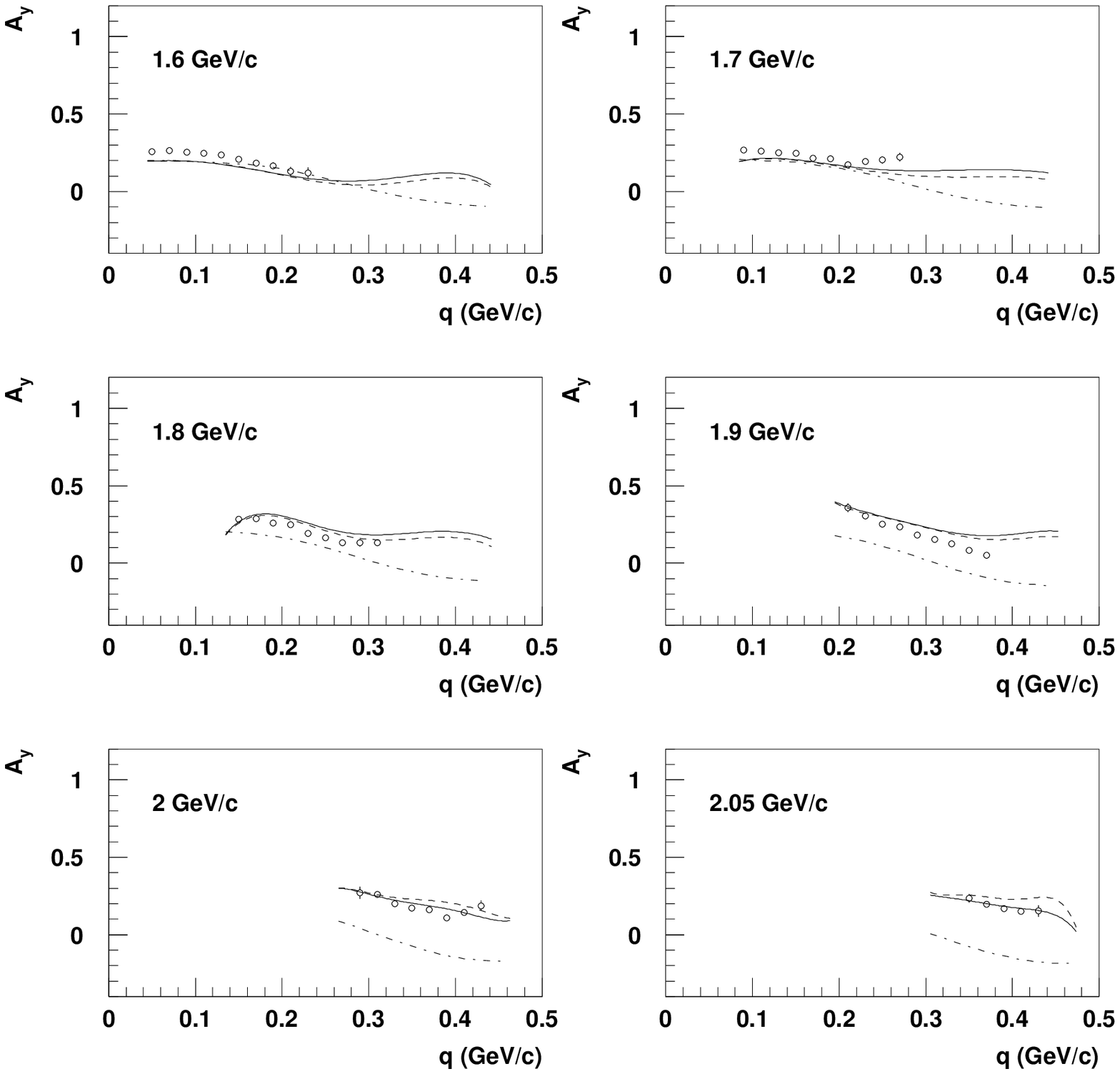,height=16cm}
\end{center}
\caption{The vector analyzing power $A_{y}$ for the HES. 
Same notations as in Fig. \ref{fig:Ayy}. }
\label{fig:Ay}
\end{figure}

\begin{figure}
\begin{center}
\epsfig{file=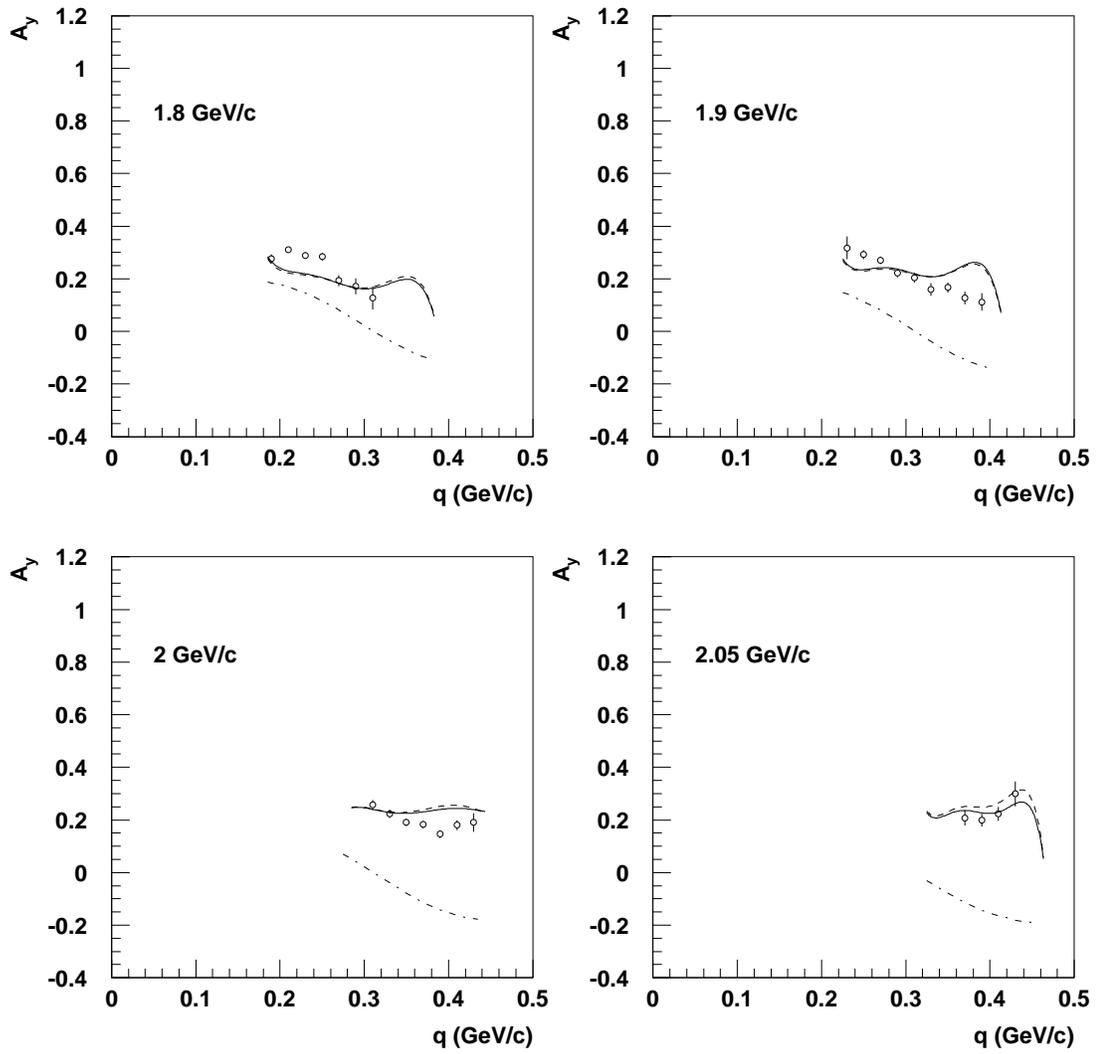,height=16cm}
\end{center}
\caption{The vector analyzing power $A_{y}$ for the LES. 
Same notations as in Fig. \ref{fig:Ayy}. }
\label{fig:leAy}
\end{figure}

The vector analyzing power $A_{y}$ results, 
shown in  Figs.\ref{fig:Ay} and \ref{fig:leAy}, 
exhibit the similar tendency: significant 
correction to the IA above 0.25 GeV/c by the full  calculations. 
The surprisingly good coincidence, taking in mind the very small statistical
errors of the measured values, of full
calculations with the experimental points is achieved  
at the all SPES-4 settings for
the both high and low energy solutions.

\begin{figure}
\begin{center}
\epsfig{file=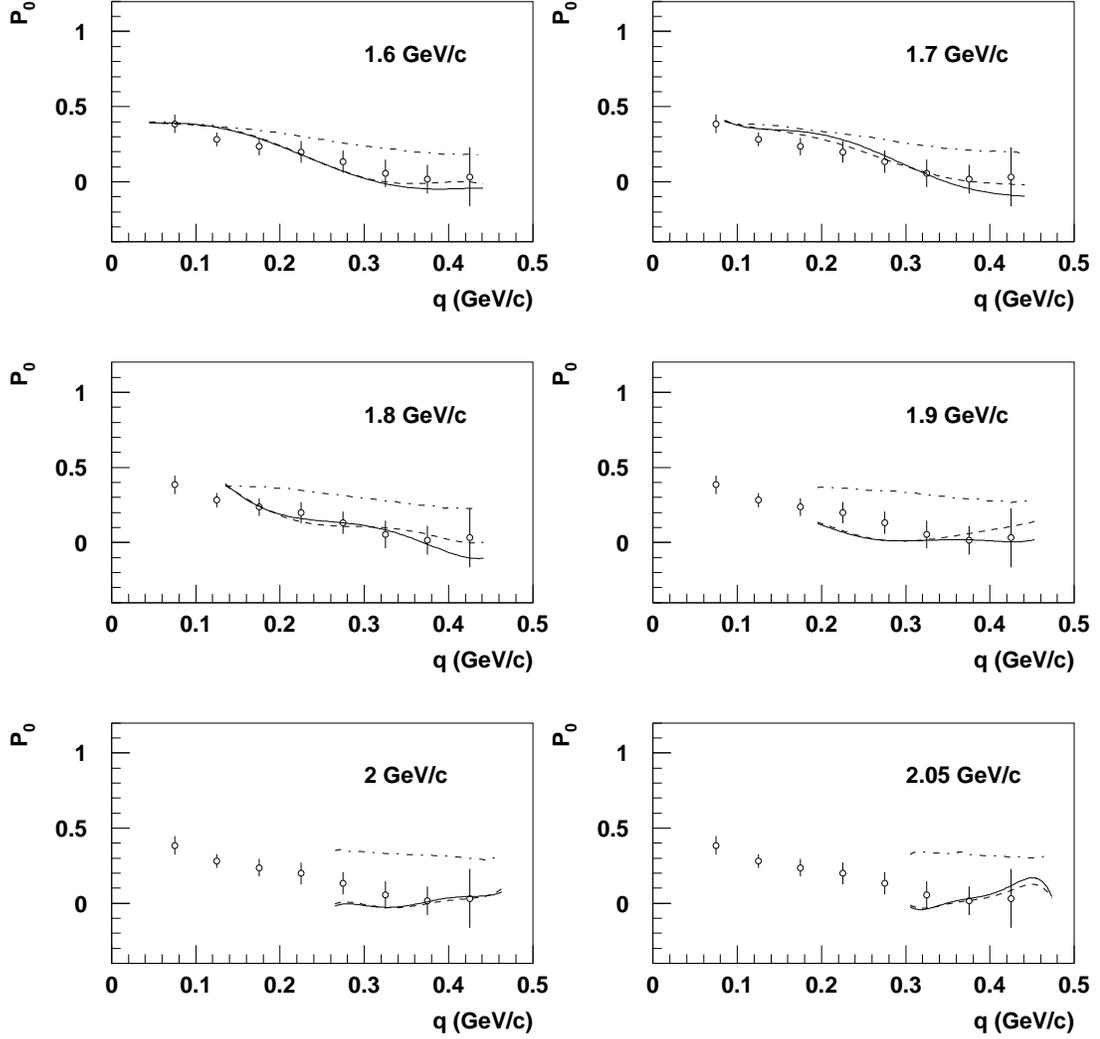,height=16cm}
\end{center}
\caption{The forward proton polarization $P_0$ for the HES. 
The notation of the curves are 
the same as in Fig. \ref{fig:Ayy}. The calculations are done consistently as for 
$A_y$ and $A_{yy}$ for each setting of the spectrometer while the data are
summed.}  
\label{fig:P0}
\end{figure}

\begin{figure}
\begin{center}
\epsfig{file=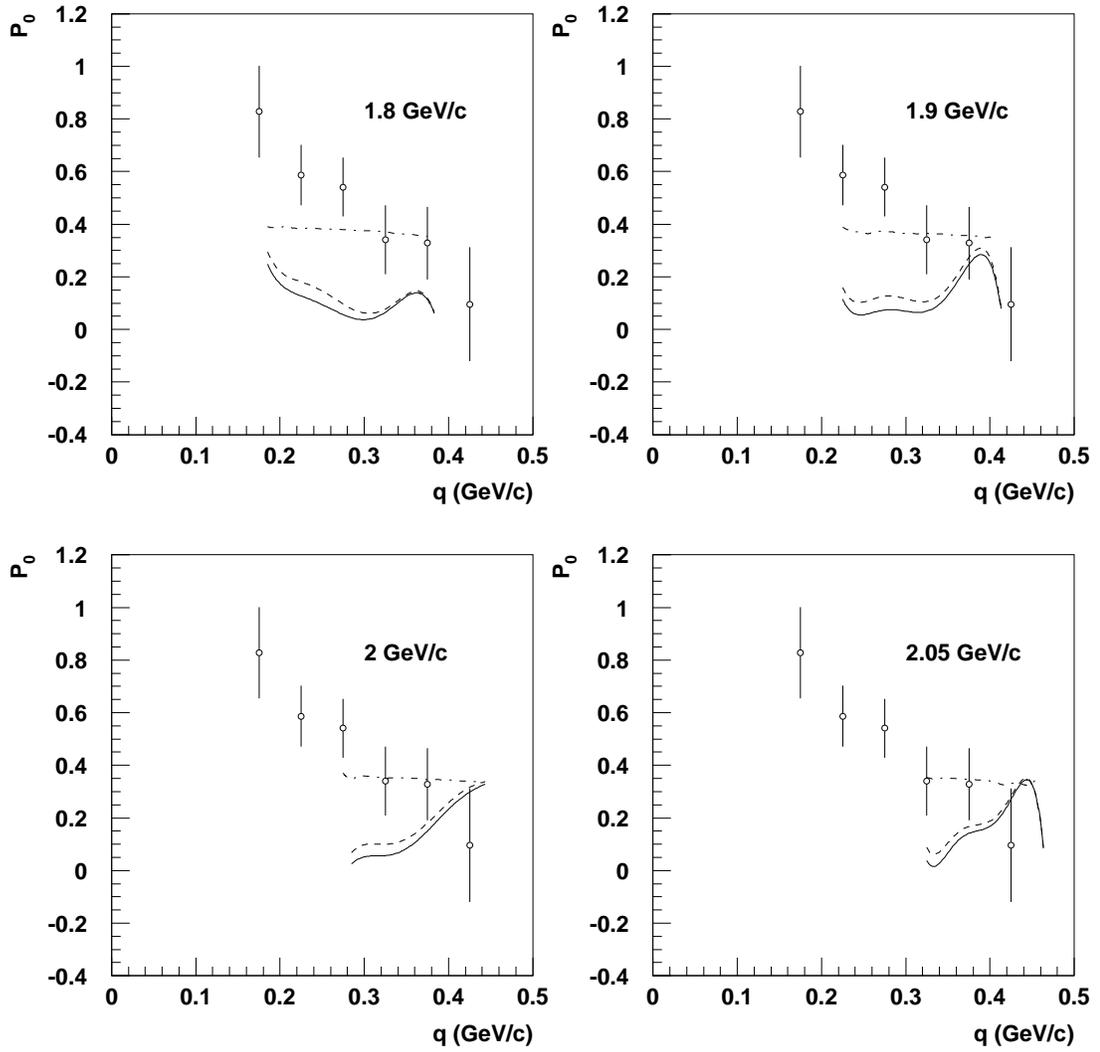,height=16cm}
\end{center}
\caption{The forward proton polarization $P_0$ for the LES. 
The notation of the curves are 
the same as in Fig. \ref{fig:Ayy}. }
\label{fig:leP0}
\end{figure}

\begin{figure} 
\begin{center}
\epsfig{file=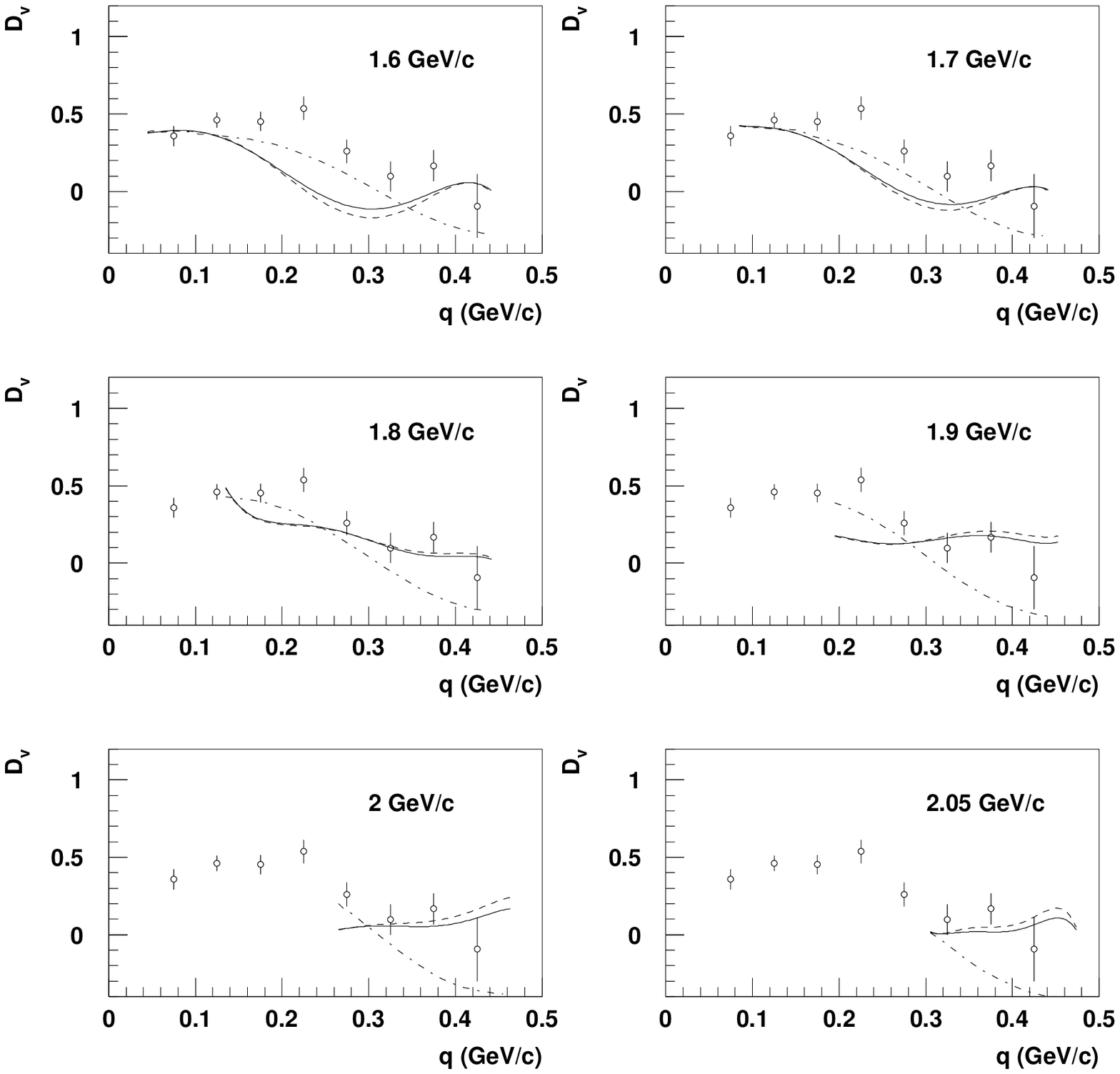,height=8cm} \epsfig{file=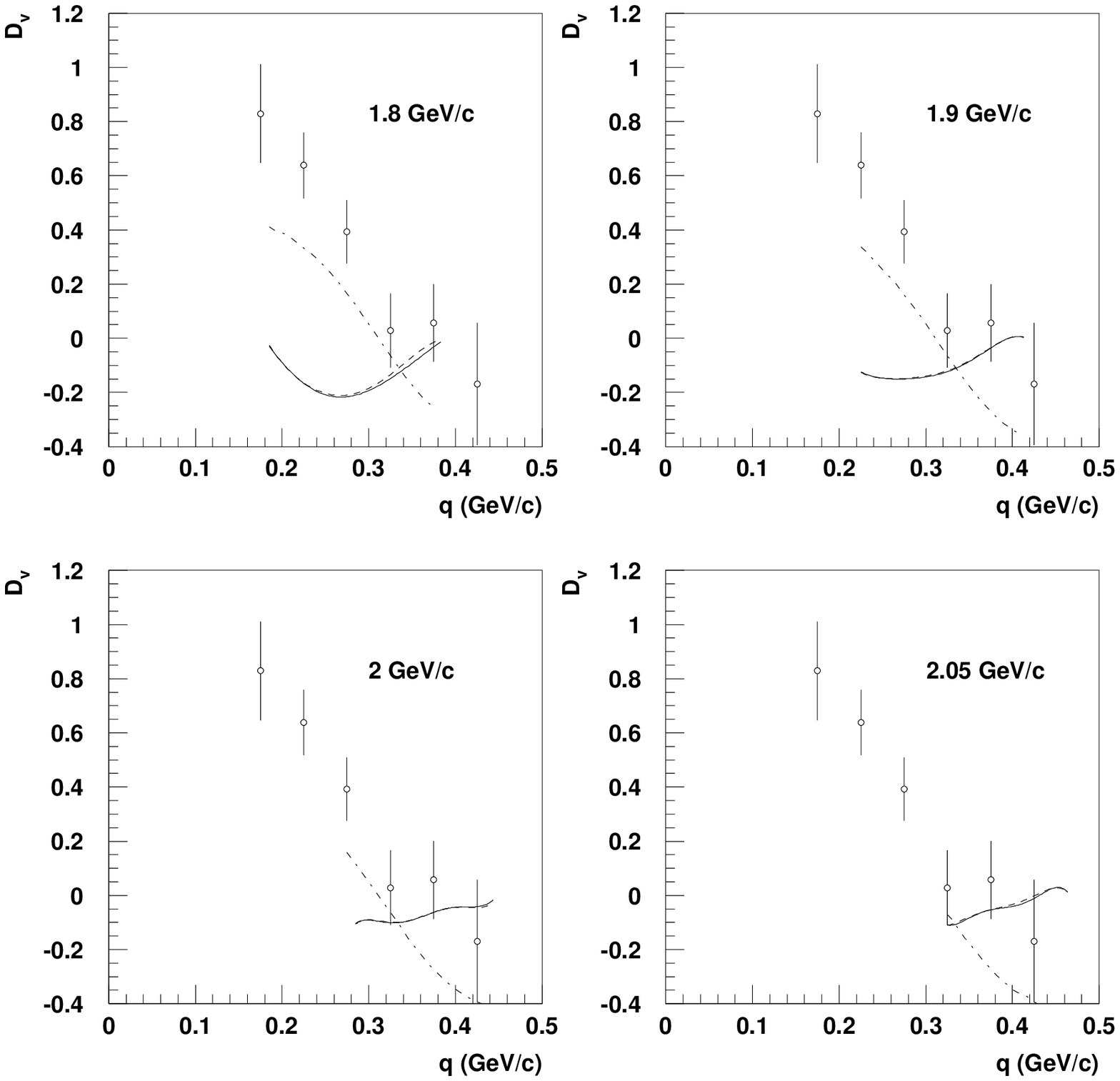,height=8cm}
\end{center}
\caption{The depolarization parameter  $D_v$ for the HES (right picture) 
and LES (left picture) . Same notations as in Fig.  
\ref{fig:Ayy}.}
\label{fig:Dv}
\end{figure}

For the polarization of the fast proton  the data are averaged over the SPES-4
settings to obtain more statistics. However the calculations were
performed for each setting.
The experimental data and the calculations are presented in the terms of
the proton polarization $P_0$  for the unpolarized 
deuteron and the depolarization parameters $D_v$ 
(see the eq.(\ref{eq:vec pol d  p1y})) and are shown 
in  Figs.\ref{fig:P0}-\ref{fig:leP0} 
 and \ref{fig:Dv}, respectively. For HES a good description is
obtained for the $P_0$ with the full diagram calculations, 
whereas it is not the case for LES. For the $D_v$, 
the IA is even closer to the data than the
full calculations. However, considering the large error bars for the $D_v$, one
can say that there is not decisive discrimination between the both calculations.

\begin{figure}
\begin{center}
\psfig{figure=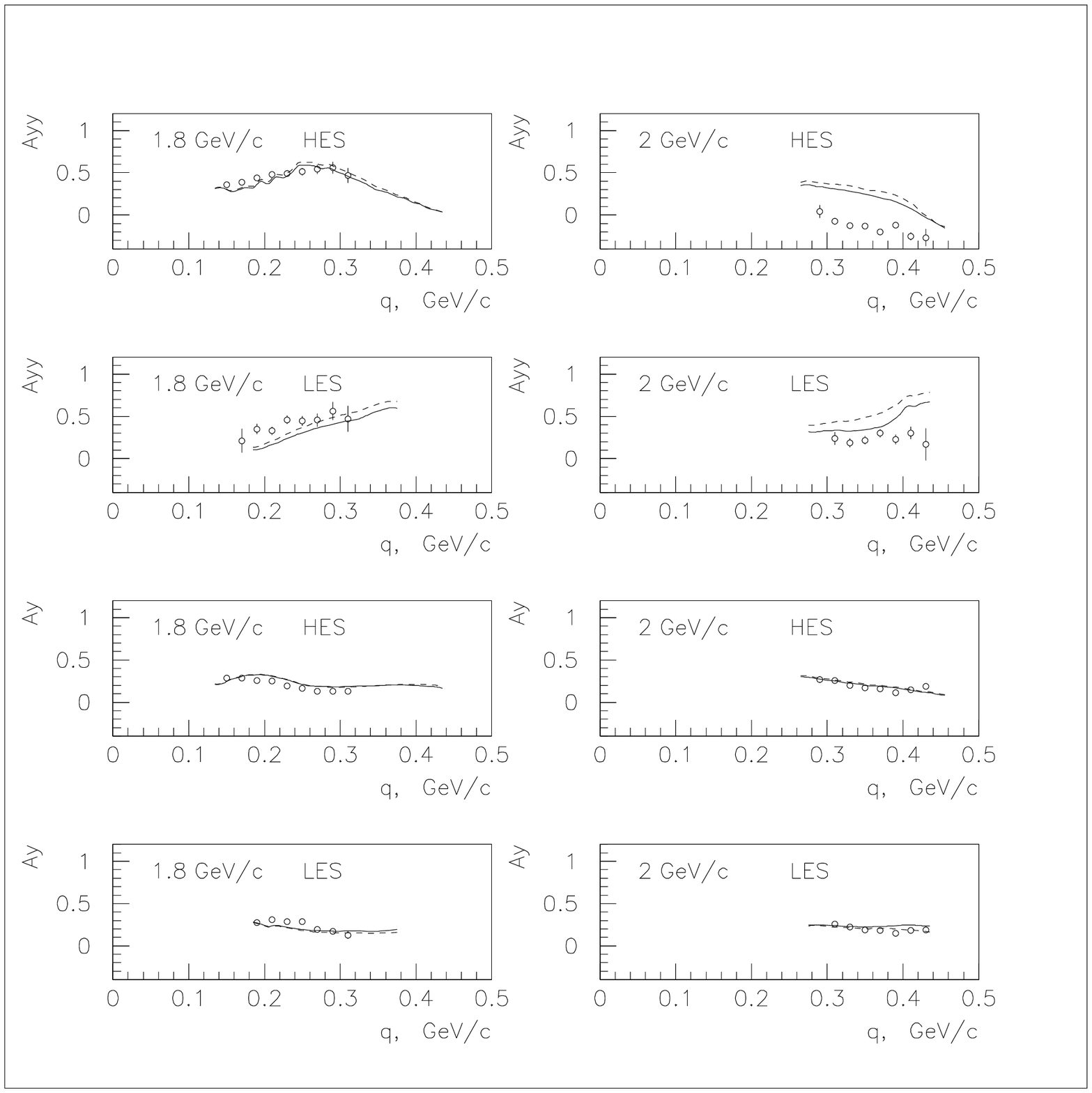,bbllx=20pt,bblly=145pt,bburx=575pt,bbury=700pt,width=16cm}
\end{center}
\caption{Comparison of the Bonn(full line) and Paris(dashed line) deuteron wave 
functions predictions} 
\label{fig:BoPaCom}
\end{figure}

The above results were obtained with the Bonn deuteron wave 
function \cite{Bonn}. Returning to the initial idea of this 
experiment to discriminate between the different deuteron wave functions we 
have tried also widely used Paris deuteron wave function  
\cite{Paris}. In Fig. \ref{fig:BoPaCom} the results of the
calculations for the both wave functions at  1.8 and 2.0 GeV/c
settings of the SPES-4 are shown. It is seen that the dependence on the 
deuteron wave function of the full calculations is  weak.

To conclude we would like to remind that the main task of this experiment
was the investigation of the deuteron structure in the high internal  
momentum region. The polarization data were thought to bring the information
on the $S$- and $D$-component of the deuteron wave function, 
which would be possible only if
the IA would dominate in the reaction mechanism. The underlying idea of this 
experiment as well as  the analogous ones is that  the deviations from 
the IA, corrected by other presumingly small contributing conventional mechanisms, 
could be interpreted as the revealing of the quark degrees of freedom.  In fact,
the polarization parameters, measured at first time in this exclusive experiment 
with a large accuracy and in a large amount, deviate strongly from the IA at the 
high internal momenta.  However, the estimations of other 
traditional reaction mechanisms, 
the nucleon-nucleon double-scattering being the most significant,  
were tried and have shown a qualitative agreement to the referred experimental
data. This agreement between the data and the theory is thought to imply that the
possible new degrees of freedom in the deuteron structure contribute much 
smaller than the hadronic degrees. Still definite perfection of the model is 
necessary for the adequate description of the polarized deuteron break-up 
data. Investigation of the short range deuteron wave function, if it is at all 
possible in the deuteron break-up experiments with the hadron probes, will be, 
by our opinion, the further step  only  after all
discrepancies in theoretical description of polarization observables are 
overcome.

\section*{Acknowledgements}

The author thanks S.Belostotski, A.Boudard,  J.-M.Laget and V.Nikulin 
for fruitful discussions and cooperation during performing this work. He is also  
very grateful to  A.Boudard for the warm
hospitality at  Service de Physique Nucleaire, 
Centre d'Etudes de Saclay, where the main part of this work was done. 

\vspace{1cm}
\section*{Appendix. M-functions}

Let us consider the process with the $n_f$ and $n_i$ spinor particles in the final
and initial channels, respectively. We will denote the corresponding
amplitude with the indices as \[ {{\rm M}^{\sigma_{1_f} \ldots \sigma_{n_f}}}_
{\sigma_{1_i} \ldots \sigma_{n_i}}\;.\]
Let $v_i$ be the boosts of i-th spinor particle moving with
the velocity $V_i$. Then the connection between the
$S$-matrix and M-function looks like
\begin{equation}
S=v_{1_f}^{-1}\otimes \ldots \otimes v_{n_f}^{-1}\;{\rm M}\;
v_{1_i} \otimes \ldots \otimes v_{n_i}\;.
\label{eq:S=invvfMvi}
\end{equation}
The boost matrix  $v^a_b$  from the rest to the 4-velocity 
$(V_0,{\bf k}\sqrt{V_0^2-1})$, where ${\bf k}$ is the unit vector along 
the momentum, is equal to
\begin{equation}
v^a_b=\left(\begin{array}{cc}
v_0+v_3 &  v_1-iv_2 \\
v_1+iv_2 &  v_0-v_3  \end{array} \right)\;,
\label{eq:boost}
\end{equation}
where 
\[v=\frac{1}{\sqrt{2}}(\sqrt{V_0+1},{\bf k}\sqrt{V_0-1})\;.\]

With respect to Lorentz transformations,  represented by the 
unimodular matrices $z$
\[z^a_b=\left(\begin{array}{cc}
z_0+iz_3 &  iz_1+z_2 \\
iz_1-z_2 &  z_0-iz_3  \end{array} \right)\;,\]
where $z_{0,1,2,3}$ are the complex numbers such that $z_0^2+z_1^2+z_2^2+z_3^2=1$,
the M-functions behave as the operators in the direct product of spinor spaces
\begin{equation}
{\rm M} \rightarrow z \otimes \ldots \otimes z\;{\rm M}\;
z^{-1}\otimes \ldots \otimes z^{-1}\;.
\label{eq:Ml=zMinvz}
\end{equation}
The simplicity of the transformation (\ref{eq:Ml=zMinvz})
is the main advantage of  the
M-functions. It allows to build the M-functions  from
products $\tilde{V}_iV_j$ of two by two  matrices  
$\tilde{V}_i \equiv V_i^\mu \tilde{ \sigma_\mu}$ and
$V_j \equiv V_j^\mu  \sigma_\mu$, where  $V^\mu _i$ are the
four-velocities of the scattered spinor particles completed by the 
four-velocity $V$ of the whole system in the arbitrary frame. The matrices
 $\sigma^\mu=(1, \vec{\sigma})$ and  $\tilde{\sigma}^\mu=(1, -\vec{\sigma})$, 
where $ \vec{\sigma}$ are the standard Pauli matrices, have the different
position of spinor indices specified in the following way
\[\sigma^\mu \rightarrow \sigma^\mu_{\bar{a}b}\;,\;\;
\tilde{\sigma}^\mu \rightarrow {\sigma^\mu}^{a\bar{b}}\;.\] 

The simplest M-functions are easy to derive using the Dirac bispinors
in  the Weyl representation in the M-function form
\[{u^\alpha}_b=\sqrt{m}
\left(\begin{array}{c}
e^a_b \\
V_{\bar{a} b}
\end{array}\right) \;,\;\;\;
{\bar{u}^b}_\alpha=\sqrt{m}
\left(\begin{array}{cc}
e^b_a   & V^{b\bar{a}}
\end{array}\right)\;.\]
Here $\alpha$ is the bispinor index, 
$V_{\bar{a} b}=V_\mu \sigma^\mu_{\bar{a}b}\;,
V^{b\bar{a}}=V_\mu {\sigma^\mu}^{b\bar{a}}$ and $V_\mu$ is the four-
velocity of the particle. 
Let us consider for example the $\pi$NN vertex. 
In the Weyl representation the $\gamma_5$-
matrix looks like \[\gamma_5 =\left(\begin{array}{cc} e & 0 \\
0 & -e \end{array}\right)\] and the amplitude of the transition between real
nucleons with the production of the pion is equal to
\[ {\rm M}^a_b(p_f,q;p_i)=
g_\pi {\bar{u}^a}_\alpha {\gamma_5}^\alpha _\beta {u^\beta}_b =
g_\pi m(e^a_b-V_f^{a\bar{c}}{V_i}_{\bar{c}b})\;,\]
or in the index-less form
\begin{equation}
{\rm M}(p_f,q;p_i)=g_\pi \bar{u}_f \gamma_5 u_i =
g_\pi m(e-\tilde{V}_f V_i)\;.
\label{eq:indless pseuds vertex}
\end{equation}

The matrices  $\tilde{V}_i$ and $V_i$, where $V^\mu _i$ is the
four-velocity of the i-th particle, serve as the metric tensors for the spin 
index of this particle in Stapp formalism when performing the contraction over 
this index (traces, successive processes).  So, in contrast to the simple
expression of the cross section via the $S$-matrix elements
\[\sigma =\frac{1}{2^{n_i}}Tr(S S^\dagger)\;,\] 
the same via the M-functions looks like
\[\sigma =\frac{1}{2^{n_i}} {\rm M}^{\sigma_{1_f} \ldots \sigma_{n_f}}_
{\sigma_{1_i} \ldots \sigma_{n_i}} V^{\sigma_{1_i} \bar{\sigma}_{1_i}}_{1_i}
\ldots V^{\sigma_{n_i} \bar{\sigma}_{n_i}}_{n_i}
\bar{{\rm M}}^{\bar{\sigma}_{1_f} \ldots \bar{\sigma}_{n_f}}_
{\bar{\sigma}_{1_i} \ldots \bar{\sigma}_{n_i}}
{V_{1_f}}_{\bar{\sigma}_{1_f} \sigma_{1_f}} \ldots 
{V_{n_f}}_{\bar{\sigma}_{n_f} \sigma_{n_f}}\;, \]
or in the index-less form
\begin{equation}
\sigma =\frac{1}{2^{n_i}} Tr({\rm M}\; \tilde{V}_{1_i} \otimes \ldots 
\otimes \tilde{V}_{n_i} \; {\rm M}^\dagger 
V_{1_f} \otimes \ldots \otimes V_{n_f})\;. 
\label{eq:M funct norm}
\end{equation}

At last let us give one of the basis of the M-functions for the NN scattering. 
It is convenient to use such  functions $b_i$ in the development
(\ref{eq:M func devel}), that the cross section is equal to 
$\sum_i|g_i|^2$. From the eq.(\ref{eq:M funct norm}) follows that 
for this purpose one should  build the basis orthonormalized with
respect to the scalar product
\[({\rm M}_i,{\rm M}_j) \equiv \frac{Sp({\rm M}_i\tilde{V}_0 \otimes \tilde{V}_v
{\rm M}_j^\dagger V_1 \otimes V_2)}{4}\;.\]
The eqs.(\ref{eq:bi}-\ref{eq:norms Mi}) present such a basis \cite{Grebenyuk1989}
\begin{equation} 
b_i \equiv \frac{{\rm M}_i}{\sqrt{\|{\rm M}_i\|^2}}\;,i=1 \ldots 6\;,
\label{eq:bi}
\end{equation}
where 
\begin{eqnarray}
{\rm M}_1=&(e-\tilde{V}_1 V_v) \otimes (e-\tilde{V}_2 V_0)\;.
\nonumber  \\
{\rm M}_2=&(\tilde{V}_1 V-\tilde{V} V_v) \otimes
(\tilde{V}_2 V-\tilde{V} V_0)\;.
\nonumber  \\
{\rm M}_3=&(e+\tilde{V}_1 V_v) \otimes (e+\tilde{V}_2 V_0)\;,
\label{eq:Mi}  \\
{\rm M}_4=&
(\alpha^{1v}(e+\tilde{V}_1 V_v)-(\tilde{V}_1 V+\tilde{V} V_v))
\otimes
(\alpha^{20}(e+\tilde{V}_2 V_0)-(\tilde{V}_2 V+\tilde{V} V_0))\;,
\nonumber  \\
{\rm M}_5=&
(e+\tilde{V}_1 V_v) \otimes
(\alpha^{20}(e+\tilde{V}_2 V_0)-(\tilde{V}_2 V+\tilde{V} V_0))\;,
\nonumber  \\
{\rm M}_6=&
(\alpha^{1v}(e+\tilde{V}_1 V_v)-(\tilde{V}_1 V+\tilde{V} V_v))
\otimes  (e+\tilde{V}_2 V_0)\;, \nonumber
\end{eqnarray}
with
\begin{equation}
\alpha^{ij} \equiv 2\frac{(V,V_i+V_j)}{(V_i+V_j)^2}\;,
\label{eq:NN alpha definition}
\end{equation}
and 
\begin{eqnarray}
\|{\rm M}_1\|^2= & 4\left[1-(V_1,V_v)\right]\left[1-(V_2,V_0)\right]
                                                        \nonumber  \\
\|{\rm M}_2\|^2= & 4\left[1+(V_1,V_v)-2(V_1,V)(V_v,V)\right]
\left[1+(V_2,V_0)-2(V_2,V)(V_0,V)\right]              \nonumber  \\
\|{\rm M}_3\|^2= & 4\left[1+(V_1,V_v)\right]\left[1+(V_2,V_0)\right]
                                                        \nonumber  \\
\|{\rm M}_4\|^2= & \frac{\|{\rm M}_5\|^2\|{\rm M}_6\|^2}{\|{\rm M}_3\|^2}
                                             \label{eq:norms Mi}  \\
\|{\rm M}_5\|^2= & 4\frac{1+(V_1,V_v)}{1+(V_2,V_0)}
\left\{ \left[1-(V,V_2)^2\right]\left[1-(V,V_0)^2\right]-
\left[(V_2,V_0)-(V,V_2)(V,V_0)\right]^2 \right\}         \nonumber \\
\|{\rm M}_6\|^2= & 4\frac{1+(V_2,V_0)}{1+(V_1,V_v)}
\left\{ \left[1-(V,V_1)^2\right]\left[1-(V,V_v)^2\right]-
\left[(V_1,V_v)-(V,V_1)(V,V_v)\right]^2 \right\}\;.      \nonumber
\end{eqnarray}
The $V_i$ in the above equations are the four-velocities of the i-th particle,
$V$ being the four-velocity of the whole NN c.m.  system.

%\bibliographystyle{unsrt}
%\bibliography{finsac}
\end{document}